\documentclass[usenatbib, usedcolumn]{mn2e}
\usepackage{lscape}

\title{Optimal photometry for colour-magnitude diagrams and its application to
NGC 2547}
\author[Naylor et al.]
       {Tim Naylor$^{1,2}$, E.J. Totten$^2$, R.D. Jeffries$^2$, 
        M. Pozzo$^{2,3}$ C.R. Devey$^2$ and \newauthor S.A. Thompson$^2$
\\
$^1$School of Physics, University of Exeter, Stocker Road, Exeter EX4 4QL\\
$^2$Department of Physics, Keele University, Staffordshire, ST5 5BG \\
$^3$Imperial College of Science, Technology and Medicine, 
Blackett Laboratory, Prince Consort Road, London SW7 2BW 
}
\date{}

\pagerange{\pageref{firstpage}--\pageref{lastpage}}
\pubyear{}

\begin{document}

\maketitle

\label{firstpage}

\begin{abstract}
We have developed the techniques required to use the optimal photometry
algorithm of \cite{p1} to create colour-magnitude diagrams with well
defined completeness functions.
To achieve this we first demonstrate that the optimal extraction is 
insensitive to uncertainties in the star's measured position.
We then show how to correct the optimally extracted fluxes such that they
correspond to those measured in a large aperture, so aperture photometry
of standard stars can be used to place the measurements on a standard system.
The technique simultaneously removes the effects of a position dependent 
point-spread function.
Finally we develop a method called ``ghosting'', which 
calculates the completeness corrections in the absence of
an accurate description of the point spread function.

We apply these techniques to the young cluster NGC 2547 (=C0809-491), and 
use an X-ray selected sample to find an age of 20-35Myr and an intrinsic 
distance modulus of 8.00-8.15 magnitudes.
We use these isochrones to select members from our photometric surveys.
Our derived luminosity function shows a well defined Wielen dip, making
NGC 2547 the youngest cluster in which such a feature has been observed. 
Our derived mass function spans the range 0.1-6$M_\odot$ and is similar to 
that for the field and the older, 
more massive clusters M35 and the Pleiades, supporting the idea of a 
universal initial mass function.  

\end{abstract}


\section{INTRODUCTION}
\label{intro}

\cite{p1} presented a new method of carrying out 
photometry for digital images, weighting each pixel according to its 
signal-to-noise ratio.
Aside from the obvious improvement in signal-to-noise ratio, this ``optimal 
photometry'' offers over traditional aperture methods, it also provides 
robust error bar estimation, freedom from bias incurred by mis-estimating 
the point spread function (PSF), and ease of automation.
The possibility of such a scheme was first suggested by \cite{sa}
\citep[we were unaware of this work when we published][]{p1}, although
they did not apply fully optimal masks, or realize the full importance
of using the same weighting for all stars.
We originally developed optimal photometry for differential time-series 
work, and in \cite{p1} explicitly ignored the problems that 
would be posed were it applied to colour-magnitude diagrams (CMDs).
Since other groups are now beginning to utilize optimal photometry, and it
is becoming more widely available
\citep[see, for example, the Starlink release][]{photom}, it would 
seem timely to rectify this omission.
Motivated by a scientific desire to obtain the most precise photometry 
possible from data on young stellar clusters and associations, along with
a reliable estimate of the fraction of stars detected at a given flux level,
we have now developed the techniques required.
In this paper we use them to derive a mass function for the 
young open cluster NGC2547 (= C0809-491).
However, it is important to emphasize that the algorithms have been
developed whilst we have been working on datasets from a range of
telescopes, instruments and targets (including $J,H,K$ IR data).
The first result to be published was our study of low-mass stars around
$\gamma^2$ Velorum \citep{monica}, and more publications are in
preparation \citep{totten00,naylor01}.

This work should be seen in the context of a host of other observational 
projects currently underway, and data reduction programmes now available. 
The most illuminating comparisons are with
projects which, like ours, concentrate on colour-magnitude diagrams.
These include the CFHT Cluster Survey \citep{kalirai01}, the WIYN Open 
Cluster Survey \citep{hippel98} and the ESO pre-FLAMES survey 
\citep{pre-flames, eis}.
Not only do we share data reduction aims with these surveys, but our 
observational techniques are forced to be broadly similar (though there are
important differences in detail). 
In contrast the very large surveys often have specialized data acquisition 
techniques, whether they survey a large area of sky, such as 
the Sloan Digital Sky Survey \citep{sloan}, DENIS \citep{denis} and 2MASS; 
or whether they
aim to detect variability such as OGLE \citep{ogle} and MACHO \citep{MACHO}.
Algorithmically, there are two main differences between all these surveys and 
our work.
The first is that we make final photometric measurements using optimal 
extraction.
In contrast most of the other projects use some type of profile fitting.
This is often either background determination followed by 
empirical profile fitting based on DAOPHOT \citep{daophot}, or simultaneous
background and analytical profile fitting, based on SExtractor 
\citep{sextractor}.
The second major algorithmic difference is that rather than adding images of 
the same field together, and then carrying out the photometry, we have
chosen to extract photometric measurements from each image, and then
combine the measurements.
Our reasons for preferring this method are outlined in Section \ref{combine_im}.

\subsection{NGC 2547}

The open cluster NGC2547 (= C0809-491) lies at Galactic co-ordinates
$l=264.45^\circ$ 
$b=-8.53^\circ$ ($\alpha$ = 08 10 25.7, $\delta$ = -49 10 03; J2000).
\cite{claria82} placed it at a distance of 450pc
with a reddening of $E(B-V)=0.06$ and an age of 57Myr.
This age is close to that of $\alpha$ Per ($\simeq$50Myr) and
\cite{jt98} sought to use it as a comparison to ascertain if age was the 
main driver of cluster properties such as coronal activity among
low-mass stars.
In doing so they discovered that fitting isochrones to the 
$V$ vs. $V-I_c$ colour-magnitude diagram for the low-mass stars, implied
an age of only 14$\pm$4Myr.
Furthermore, they discovered very little scatter of the stars about the 
isochrones, beyond that attributable to binary systems and photometric errors.
It is this latter property which attracts us to NGC 2547 as a testbed for our
new photometry techniques.
Reproducing the narrow pre-main-sequence, where the stars are widely dispersed
over several CCD fields is a stringent test of both aperture correction and
correctly allowing for the spatial dependence of the point spread function.
Furthermore, as we have already reduced the same dataset using aperture 
photometry, a direct comparison is possible.
Finally, the data were clearly suitable to yield a further scientific result,
namely the cluster mass function. NGC 2547 is sufficiently young that
it is unlikely to have suffered equipartition and dynamical evaporation
of its low-mass members, so the present day mass function may be
representative of the {\em initial} mass function. In addition, it
seems likely that NGC 2547 is much less massive than previously well studied
open clusters like the Pleiades and NGC 2516 (it has far fewer
early-type stars). We therefore have the opportunity to see whether the mass
function varies with total cluster mass or is scale invariant.

\subsection{An outline of the paper}

To help the reader keep sight of how the various parts of the reduction
process are linked, we will outline the reduction process, and
hence the structure of the paper.
We begin by outlining the observations in Section \ref{obs}.
There are multiple images in each band for all of our fields.
Optimal 
photometry forces us to measure each star separately on each image, and 
then combine the results (Section \ref{combine_im}).
To do so we must consider the effects of mis-centering the mask on the 
accuracy of our results, and we derive this for optimal
photometry in Sections \ref{mask-err} and \ref{cent-err}.
We have found there are significant advantages to
be gained by tying our source detection algorithm to our
photometry technique, so the former is outlined in Section \ref{detect}.

At the same time as detecting the stars, we find their positions.
This involves the only significant difference between the extraction code 
presented in \cite{p1} and that used here.
For largely historical reasons the old version of the code determined the 
position of the star by fitting a Gaussian to a region 50 pixels square 
centered on the approximate position of the star.
We now use a box whose side is twice the full width at half maximum (FWHM) 
of the star.
In addition to a large increase in speed, this change 
has the considerable advantage that it is much less likely there will be
a second star in the region (such a star then has to be fitted).
Indeed, if there is significant flux from a second star within this box, then
the optimal photometry itself will be compromised, and the star will also
have been flagged as ``non-stellar'' (see Section \ref{detect}).
We are therefore unconcerned if the position is slightly inaccurate, and so
we can now always limit ourselves to fitting a single Gaussian.
After the positions have been determined, optimal photometry is then carried
out as described in \cite{p1}.

We then discuss the most serious problem we encountered in creating
colour-magnitude diagrams.
We anticipated that we would have to carry out the equivalent of
aperture correction, which we call profile correction (Section 
\ref{prof_corr}).
What we did not foresee is that the correction would be a strong function 
of position, but as we show in Section \ref{var}, dealing with this 
turns out to be a tractable problem.

After combining measurements of the same star from different images (Section
\ref{comb_meas}) we obtain astrometric positions for the stars  
(Section \ref{astrom}).
We next apply our photometric solution and compare our results for NGC2547
with those in the literature (Sections \ref{fcal} and \ref{cats}).
We then fit isochrones to the X-ray selected members to determine the cluster's
age and distance (Section \ref{age}).
The fitted isochrones allow us to select cluster members (Section \ref{cands}),
but of course we cannot use this to derive a mass function until we know
how complete the catalogue is.
Thus Section \ref{ghosts} presents our method of deriving the completeness
correction.
The reason we cannot use standard techniques for this is that they
require an accurate knowledge of the point spread function, thus we have
developed a technique we call ``ghosting''.
We can then derive the cluster luminosity and mass functions (Sections
\ref{lf} and \ref{func} respectively).
The resulting mass function is similar to other clusters, giving us confidence
in our techniques.
Finally, in Section \ref{conc} we present our conclusions and recommendations.

\section{OBSERVATIONS}
\label{obs}

The observations were obtained in 1996 March using a $2048^2$ CCD mounted 
at the Cassegrain focus of the CTIO 0.9-m telescope.
This yielded a field
of view of approximately $13 \times 13 $ arcminutes.
Full details are given in \cite{jt98}, but the important point
for this work is that the data consisted of two $BVI_c$ surveys, both
centered on the cluster.
A deep survey of 4 CCD fields covered the central regions of the cluster,
whilst a wide survey of 9 fields covered the same central regions, and a
contiguous outer area.  
With one exception, the wide survey consisted of both a long and a short 
exposure for each field in each band, whilst the deep survey was of 3 equal
length exposures in each field in each band.
The exposure times are given in Table 1.

Common practice is to move the telescope between each exposure in a series,
such as our deep fields.
This has the effect of moving the bad columns around the sky, so that all
stars within the survey ``box'' will be measured in some frames.
In addition it breaks up any residual flatfield or night sky emission line 
fringing effects.
However, it also gives a non-uniform limiting magnitude to the final catalogue.
For our observations the telescope was not moved, yielding a more uniform
limiting magnitude, but meaning that we have not covered the entire area of
sky within our survey box.
Which strategy is best depends on the primary purpose of the observations.
For deriving a mass function the shape of the area surveyed is unimportant,
but simplicity of the completeness correction is valuable.
In addition our limiting magnitude lies well above any flatfield effects, and
so our strategy of not moving the telescope is well suited to the experiment
in hand.

Standard stars were observed on all nights, but as stated in
\cite{jt98}, these observations showed that only the long exposures
for the wide survey were taken in photometric conditions.  
For consistency we reduced the standard stars for this night using the
aperture techniques outlined here and in \cite{p1}.
The resulting zero points were fitted assuming linear colour terms.
For the V-band we found we had to add (in quadrature) an additional,
magnitude independent uncertainty of 0.017 magnitudes to obtain a
$\chi ^2 _\nu$ of about one, but that essentially no such increase was
required in $B-V$ or $V-I_{\rm c}$.

It should be noted that the reddest
standards observed had a $V-I_c$ of 2.7 and $B-V$ of 2.2.

\begin{table*}
 \centering
 \begin{minipage}{140mm}
  \caption{Log of Observations.}
  \begin{tabular}{@{}cccccccc@{}}
   Field    & \multicolumn{2}{c} {Field Center (J2000)} & 
 \multicolumn{3}{c} {Exposures (s)} \\
   number   & R.A.       &  Dec. & $B$ & $V$ & $I_c$ &  \\
        &  &  &  &  &  & & \\
1  & 08 10 37.4 & -49 17 40 & 3$\times$200 & 3$\times$100 & 3$\times$80   \\
2  & 08 09 33.6 & -49 17 40 & 3$\times$200 & 3$\times$100 & 3$\times$80   \\
3  & 08 10 37.5 & -49 07 10 & 3$\times$200 & 3$\times$100 & 3$\times$80   \\
4  & 08 09 33.6 & -49 07 10 & 3$\times$200 & 3$\times$100 & 3$\times$80   \\
7  & 08 09 07.7 & -49 23 30 & 30, 250 & 15, 100 & 15, 100\\
8  & 08 10 12.0 & -49 23 30 & 30, 250 & 15, 100 & 15, 100\\
9  & 08 11 16.3 & -49 23 30 & 30, 250 & 15, 100 & 15, 100\\
12 & 08 11 16.3 & -49 13 00 & 30, 250 & 15, 100 & 15, 100\\
13 & 08 10 12.0 & -49 13 00 & 2$\times$30, 250 & 100 & 2$\times$15, 100\\
14 & 08 09 07.7 & -49 13 00 & 30, 250 & 15, 100 & 15, 100\\
17 & 08 09 07.7 & -49 02 30 & 30, 250 & 15, 100 & 15, 100\\
18 & 08 10 12.0 & -49 02 30 & 30, 250 & 15, 100 & 15, 100\\
19 & 08 11 16.3 & -49 02 30 & 30, 250 & 15, 100 & 15, 100\\
\end{tabular}
\end{minipage}
\end{table*}

\section{COMBINING THE IMAGES}
\label{combine_im}

Given that one has several images of the same field, there are broadly
two ways to proceed.
The first is to simply add all the images together, and perform
photometry on the combined image.
The second is to carry out photometry on each image individually, and
then combine the results for each star.
The problem with adding several images together is that
images of poor seeing, or high sky background will compromise the
signal-to-noise ratio available from the better quality images.
Despite this, other projects aiming to create colour-magnitude diagrams
(see Section \ref{intro}) have taken this approach.
To precisely align the images they must then rebin the data into 
sub-pixels before adding the images using an algorithm such as ``drizzle''
\citep{drizzle}.
This procedure removes the statistical independence of the pixels in
the final images, complicating noise estimates.
Thus we believe the correct approach is to extract photometric measurements
for each star from each image, and then combine them by a weighted
mean.
There is another, probably more important consideration which forces
us to the same conclusion.
We find that our profile correction technique works well for individual
images, but that the scatter about the fit described in Section \ref{prof_corr}
increases if several images are added together.
We have no convincing explanation of this effect, but it gives an additional
reason for dealing with individual frames.
There are three final reasons for adopting this approach.
First, the profile correction is probably the limiting factor
at high signal-to-noise ratios, and by measuring each star in many frames we 
average out the effects of errors in its determination.
Second, stars which lie on bad pixels in one image, may not do so in other 
images.
Combining the measurements, not the images, allows one to obtain good 
photometry for these stars by simply rejecting the affected data, rather
than by resorting to an image patching algorithm.
Third, by testing for consistency between the different measurements of the
same star we can find either variable stars, or objects which have landed
on uncatalogued chip defects (see Section \ref{comb_meas}).

There are two disadvantages of using individual images, both related
to faint objects.
The first is that faint sources visible in a summed image are simply not 
detected in individual images.
Secondly, if one centroids a faint source, the resulting flux will be
biased in such a way as to be too bright \citep[see][]{p1}.
Thus measuring a faint source in several images can result in a
spectacular bias. 
The solution to these problems is to run the source detection algorithm,
and position measurement on a summed frame.
Since we are only interested in relatively bright stars (say signal-to-noise 
ratio better than 10), it does not matter that measurements of stars in this 
image do not reach the signal-to-noise ratios obtainable from combining
measurements from individual images.
The detection can be performed
at a signal-to-noise ratio of, say, three, and spurious sources removed when
the final catalogue is made.
If a lower signal-to-noise ratio cut-off were required \cite{fischer94} shows that
once the detection cell size is chosen, it is possible to calculate the
weightings required for each image to maximize the signal-to-noise ratio in the
summed image.
We can also take the positions measured in the summed frame, translate 
them into the individual frames, and fix them when performing the
photometry.
This avoids the centroid biasing in an analogous manner to the procedure 
outlined in \cite{p1} for time series work on eclipsing binaries.
The only issue we must address, is that the translation of the co-ordinates
from the summed image to the individual images must be accurate
enough not to affect the resulting photometry.
To solve this problem, we must assess how sensitive the optimal extraction 
is to mis-centering of the mask.
The one problem we have not been able to solve for the combined
images is that of bad pixels.
Practically we find we have to patch these (using a median of the 
surrounding data) before the individual frames are combined.
This will have a small effect on the astrometry, but not of course on
the photometry.

\subsection{The mask mis-centering error}
\label{mask-err}

\begin{figure}
\vspace{70mm}
\includegraphics{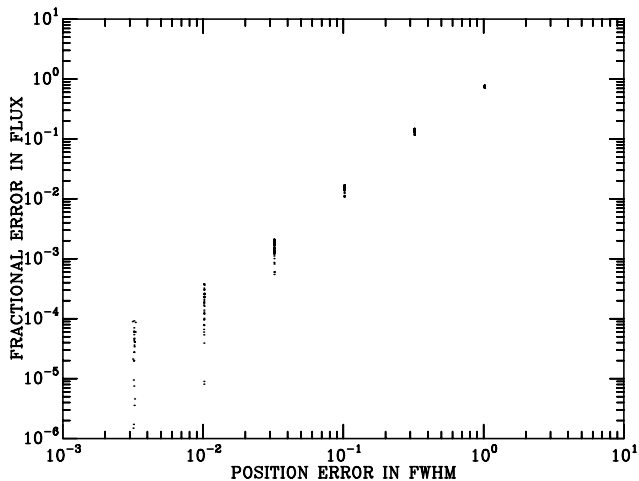}
\caption{
The error in the flux measurement as a function of the distance the
extraction mask is mis-centered.
}
\label{position}
\end{figure}

In \cite{p1} we assumed that the centering of the mask was perfect, and
did not contribute a significant error to the photometry.
However, since the optimal mask is strongly weighted towards the
center of the star, it is clear that it is more sensitive to mis-centering
errors than aperture photometry.
To assess this sensitivity we measured the flux from each of ten stars
through a series of masks that were deliberately offset from the
measured centroid.
For each star and each mis-centering distance, we made four estimates of
the flux at the four cardinal points with respect to the pixel grid.
We then divided these off-center estimates by the on-center one to
discover the fractional error the mis-centering produced.
The results are shown in Fig. \ref{position}.
The most important conclusion from this plot is that only when the centering
error approaches a tenth of a FWHM does the error in flux exceed one percent.

\subsection{The centroid error}
\label{cent-err}

\begin{figure}
\vspace{70mm}
\includegraphics{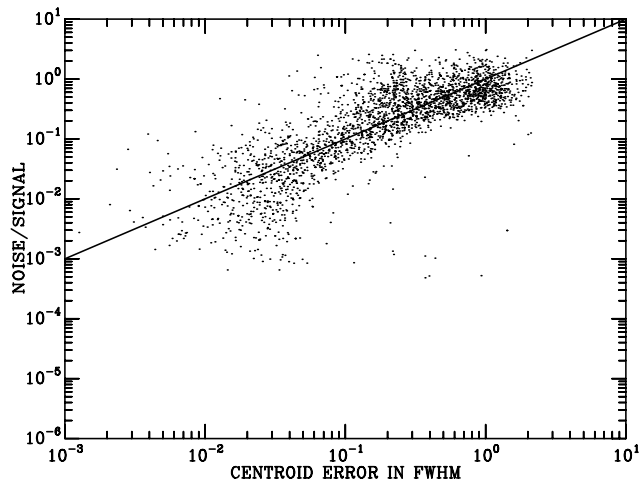}
\caption{
The error in the measurement of the position of a star as a function of
the signal-to-noise ratio.
The solid line corresponds to an error of the FWHM divided by the
signal-to-noise ratio.
}
\label{centroid}
\end{figure}

To determine if the mask centering is contributing a significant error, we
now need to estimate how accurately we are centering the mask as a function
of signal-to-noise ratio.
To assess this, we measured a set of star positions on a long and a short
exposure of the same field.
We then calculated a six-coefficient transformation of the long-frame
positions to the short-frame ones, and examined the residuals (Fig.
\ref{centroid}).
The majority of the plot follows the rule that the error in the centroid
is approximately the FWHM of the image divided by the photometric
signal-to-noise ratio \citep{king83}.
Only when the centroid error declines to about 0.04FWHM, do we see 
significant deviation from this rule, when the error becomes independent 
of the signal-to-noise ratio.
This corresponds to about a tenth of a pixel, and presumably subtle effects 
in the CCD, such as charge transfer ``coma'' and intra-pixel sensitivity
conspire to make this the smallest uncertainty one can obtain without more
sophisticated analysis \citep[e.g.][]{zacharias00}.

To combine the results of this section with that above, we next require that
the uncertainty in the photometry as a result of mis-centering is significantly
less than the photon induced noise, for a given centroid error.  
Fig. \ref{centroid} can be viewed as giving the fractional photon noise
as a function of centroid error. 
Similarly, Fig. \ref{position} gives the fractional error in flux as a result
of mis-centroiding.
Since the points in Fig. \ref{position} lie comfortably below those in Fig. 
\ref{centroid} until the signal-to-noise ratio falls to about one, 
the centroiding error is always a negligible contributor to the
signal-to-noise ratio for all practical cases.
Further, our six-coefficient solution will always provide accurate enough
positions that we can fix the positions in each frame.

\section{OBJECT DETECTION}
\label{detect}

The star detection algorithm is closely tied to the photometry technique.
We measure the sky in boxes sufficiently large that, when the sky determination
is subtracted from a star, the noise from the sky measurement is negligible
\citep[see][]{p1}, and then subtract an interpolated sky from each pixel.
Since the sky is determined by fitting a skewed Gaussian, at the same time
we find the standard deviation of the sky, excluding star pixels.
We then smooth the image using a simple top-hat filter, whose width is
the nearest value of 2N+1 to four-thirds of the seeing in pixels.
As shown in \cite{p1}, an aperture whose diameter is four-thirds of the
FWHM seeing yields the best signal-to-noise ratio for simple aperture
photometry of sky-limited objects.
Thus smoothing the image in this way is equivalent to a sliding cell detection
algorithm whose cell size is optimized for faint objects.
We can then find which pixels are $n \sigma$ above the noise, and 
use a simple north-south-east-west connectivity condition to create
a list of stars \citep[see, for example][]{irwin97}.

The main problem with such a simple technique is that stars in the wings
of other stars are lost where the flux does not fall to $n \sigma$ above
sky between
their images, as the connectivity condition assumes they are part of
the same image.
We therefore begin by thresholding the image at many tens of $\sigma$,
to create an initial list of stars.
We then re-threshold the image at a $\sigma$ level a factor of a few lower.
If a star from the previous list lies within one of the new objects, the
new object 
is discarded, otherwise it is added to the list of stars.
We repeat this process, lowering the sigma level a factor of a few at 
a time, until the required level is reached.

The final task of the star finder is to assess which objects will have their
photometry significantly affected by the presence of a nearby companion
star.
Although we refer to such objects as non-stellar, we wish to be clear that
the aim is not to find close visual binaries, merely to reject objects whose
position on the CMD is significantly affected by 
another star or stars.
We can crystallize this definition by recalling that the integral under a
two dimensional Gaussian reaches 99 percent of its full value at a radius of 
1.3 FWHM.
Thus the optimal photometry of a pair of equally bright stars will be
affected at less than the 1 percent level by the presence of a
companion 2.6 FWHM distant.
Thus, we would wish that each component is marked as ``stellar''.
Given that we wish to measure the flux through a mask of radius twice the
FWHM, the question is whether there are any deviations of the stellar profile
within that radius that, when folded through the mask, total more than some
given threshold.
Thus our criterion is that we compare the flux extracted through a mask whose
FWHM corresponds to that of the star (so roughly a normal extraction), with
that extracted using a mask whose FWHM is half that of the star.
Thus the second extraction is representative of the central regions of the
star.
We demand that this ratio is within three $\sigma$ of the median, where
the $\sigma$ used is the sum in quadrature of the uncertainty for each 
measurement and the (clipped) RMS about the median.
Of course we have no information as to whether saturated objects are
stellar or not, so we examine these objects in a short exposure, should
one have been taken.
Thus the final output from this programme is a list of stars on
which photometry is to be performed, with non-stellar objects flagged,
and a list of stellar objects, which can be used as PSF stars.

At this point it is useful to compare our choice of algorithms with
those in SExtractor \citep{sextractor}.
The comparison is particularly interesting, since SExtractor is designed to 
work well with low surface brightness galaxies, in contrast to our 
emphasis on unblended point sources. 
\cite{sextractor} considered sky estimation schemes similar to ours, and 
although acknowledging them as the best, rejected them on the grounds of 
computational cost.
Instead, they estimate the background as 2.5 times the median minus 1.5 
times the mean.
In our case, we have not found the extra compute time has a significant
impact on the overall processing time, perhaps because we only search for
sources once per field, but then carry out photometry on many images.
Both software packages use a thresholding technique to detect the stars.
SExtractor needs the flexibility to convolve the data with a range of 
possible shapes, whilst we only have to deal with point sources, hence our
choice of the simple top hat filter.
We tried varying the size of the filter, but as one might expect 
\citep[see, for example, the discussion of aperture size in][]{p1} this has
only a very small effect on the detectability of sources.
Finally, SExtractor takes great care over deblending.
We simply wish establish whether the object is sufficiently uncontaminated 
for optimal photometry, and although using a similar ``tree'' method,
can apply simpler criteria.

\section{OPTIMAL PHOTOMETRY}
\label{optphot}

Before carrying out optimal photometry, we must decide what mask to use.
In contrast to our time series work (where a PSF star could be chosen
by hand), our plethora of fields requires an automated procedure to
choose a good PSF star.
Our solution is to fit elliptical Gaussians to the brightest 49 unsaturated 
``stellar'' objects,
pick the star whose geometric mean FWHM is the median of the list,
and fix the mask parameters at the fitted values for that star.
We then perform optimal photometry of all the objects detected in the
field.

From the point at which photometry of an object is attempted, the measurement
carries with it an integer flag between zero and nine, where zero indicates 
no known problem.
Each possible problem is allocated a unique number, which is written into
the flag if it occurs; thus later problems overwrite any flag written by
earlier ones.
This process begins by writing to the flag if the detection algorithm has
determined that the star is non-stellar.  
During the optimal photometry procedure, the object's flag is changed if 
it is too close to the detector edge for any stage
of the photometry process; if the fit to the sky histogram failed; if
a saturated pixel was within the extraction mask; or if a known bad pixel
was within the mask.
%We have sometimes found it useful to flag an object if, when fitting the 
%sky histogram the skew parameter, $\beta$, in equation 17 of \cite{p1} is 
%more than 0.3. 

One of the advantages of the optimal extraction is that the error estimate
is very robust.
However, if we compare measurements of the same star in different images
(see Section \ref{comb_meas}) we find that the RMS between measurements 
never falls below $\simeq$1 percent.
Such effects are well known in CCD detectors, and probably relate to the
intra-pixel sensitivity changes.
We shall return to calculating the value of this magnitude independent
uncertainty in Section \ref{comb_meas}, but
for the moment simply assume it has a known value.

\section{Profile correction}
\label{prof_corr}

Aperture correction is used in classical CCD aperture photometry
to correct the flux measured in
the small apertures used for the majority of the objects,
to the large apertures that must be used for standard stars.
In this context the aperture correction is the fraction of the flux
which lies outside the small aperture, but inside a large one.
Obviously there must be an analogous process for optimal photometry;
comparing the flux derived from the optimal photometry, with that
in a large aperture.
For reasons explained below, we refer to this as the
profile correction.

The definition of the profile correction is slightly harder to visualize than
that of the aperture correction.
As explained in Section 2.3 of \cite{p1}, optimal extraction can be viewed as
using each pixel as an independent estimator of the total flux, where the
extraction mask gives the fraction of the flux expected in each pixel.
Combining these independent estimators in a way that optimises the 
signal-to-noise ratio is the {\it raison d'etre} of optimal extraction.
Even a single pixel, therefore, will give an estimate of the total flux, 
provided that the mask is a good description of the overall profile.
If this were the case, the profile correction would always be zero.
In general, however, the mask is only close to the actual profile, and what 
is encoded in the profile correction is the difference between the true
profile, and the extraction mask.
Thus if the extraction mask is a perfect match to the observed profile, then
the profile correction is zero; non-zero values indicate deviations
from this profile.

The difference between an aperture correction and a profile correction can be
brought into sharp focus by comparing the effects of the aperture radius
used in classical CCD photometry, and the mask clipping radius used in 
optimal photometry.
(This clipping radius, discussed in Section 3.3 of \cite{p1}, is the radius 
beyond which the optimal extraction mask weights are set to zero, to avoid 
the masks extending out to infinity.)
As the aperture radius grows, and approaches the radius used for the standard 
stars, the aperture correction tends to zero.
Conversely, as the clipping radius grows, there is no change in the profile
correction -- providing that the extraction mask is a good model of the PSF. 

\subsection{The spatially invariant case}
\label{inv}

The simplest way to perform this correction is to measure a selection
of stars with both the large aperture and the optimal technique,
and take the modal value for the difference between these measurements
as the correction.
The modal correction is chosen since some stars will have other stars
within the large aperture, thus skewing the mean away from the required
value.
In practice we find that for reasonable aperture sizes, in open cluster
fields, even the mode is systematically dragged.
We have found a more robust method is to extract sub-images centered on 
bright stars from the CCD image.
We then ``median stack'' several of these sub-images to create an observed
PSF, from which the profile correction can be calculated.
If there are other stars within the large aperture of any of the sub-images,
the median stacking process ensures their removal.

The details of this method are as follows.
We select 80 stars and measure the sky for each one (making sure the
uncertainty in the sky measurement does not contribute significantly to the
uncertainty in the large aperture flux measurement).
We then create a sky subtracted image of each star, normalized such that
the flux measured in the optimal mask is the same for each star.
If the center of each star lay at (say) the center of a pixel, one could
now create a median image, but in practice the center of each star can
be at any position within the central pixel.
We therefore begin with the brightest star, and re-sample all the other stars 
onto its pixel gridding, before creating the median stacked image.
This works well in the outer regions of the profile, but at the image 
center, where the gradient of the profile is changing most rapidly, the
interpolation tends to smooth the profile.
Within this region, therefore, we simply set the pixel values to those of
the brightest star.
We then repeat this process, creating median images based on successively
fainter stars.

It is important to note that the stellarity test has ensured that
within the region where median stacking is no longer used (a radius of
less than four-thirds of the FWHM), none of the stars chosen are
significantly contaminated by another star.
Each image should, therefore, produce a reliable profile correction.
We normally use the median of the images resulting from the 20 brightest
stars, and use the standard deviation as an estimate of the uncertainty in
the profile correction.

\subsection{The spatially variant case}
\label{var}

\begin{figure}
\vspace{70mm}
\includegraphics{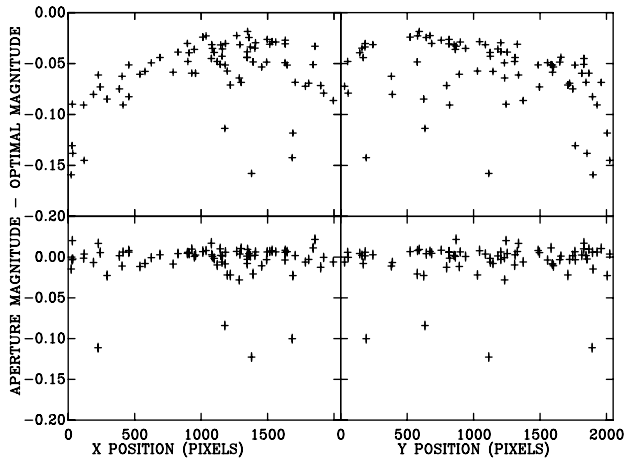}
\caption{
The profile corrections as a function of position (upper panels) and their
residuals from a two-dimensional polynomial fit which has four terms in each
direction, plus a constant (lower panels).
The error bars are dominated by the magnitude independent uncertainty 
(0.005 mags).
In the fit for the lower panels 5 data points were zero weighted, yielding a 
$\chi ^2 _\nu$ of 3.4 and an RMS of 10 milli-mags.
}
\label{apcor}
\end{figure}

If the point spread function is itself a function of position, then the 
profile correction must also be supplied as a function of position.
The obvious generalization of the above procedure is to divide the
CCD into small areas, generate a profile correction for each of them, and
then interpolate the correction.
In practice such a procedure will run out of bright stars, and so is
impractical.
Fortunately, the fields where we have encountered a spatially variant
PSF, are also sufficiently uncrowded that we can revert to using the
profile corrections calculated for individual stars as outlined at the
beginning of Section \ref{inv}.
We then fit the corrections with a low-order two-dimensional
polynomial (after adding the magnitude independent uncertainty derived
in Section \ref{comb_meas}), applying a clipping procedure to remove
those objects where the presence of a star in the large aperture gives
an anomalous correction.
This effect can be seen in Fig. \ref{apcor}, where the  
upper panels show the profile corrections as a function of X and Y 
position, and the lower panels show the residuals from the polynomial 
fit.
The objects in the lower panels which lie below the 
polynomial fit are those which have other stars within their large apertures,
and have been zero weighted.

Interestingly, we do not always obtain the best results by applying the
above method in its purest form.
Sometimes using the method outlined in Section \ref{inv} on stars taken
from the central region of the CCD, is a better way of setting the constant 
in the polynomial.
The way to decide which method is the correct one to use, is look at the stars in common 
between CCD images which overlap, and examine how closely the magnitudes
deduced from each field match.
Although we only have a limited number of datasets to test the two methods
for setting the constant, it appears that the pure polynomial method works 
best when the profile correction is a strong function of position.
Presumably this is because the profile correction varies even over the small 
area used to create the median profile.
However, the median profile method is more robust against many faint stars 
lying within the large aperture, and so works well if the profile is only
a slowly varying function of position.

For the NGC2547 $V$-band data, we found the RMS scatter between fields (before
the correction procedure described in Section \ref{fcal}) fell 
from 0.045 mags to 0.022 mags on changing from using the median profile to
the pure polynomial.
The details of how this number was derived are discussed in Section 
\ref{fcal}, however it clearly indicates that the pure polynomial method was
the correct one to use for these data.

\section{COMBINING MEASUREMENTS}
\label{comb_meas}

The data are dealt with in groups of exposures of the same field taken at
similar airmass.
The raw counts from the optimal extraction routine are first converted into
instrumental magnitudes.
For measurements with negative counts we use the modulus of the count rate,
and, if the flag is currently zero, set it to a particular value.
This procedure not only preserves the information for objects with negative 
fluxes, but also means the calibration steps involving adding a correction
in magnitude space have the correct effect.
The profile correction is then applied, the data are corrected to 
the mean airmass of the
group (using the photometric co-efficients derived in Section \ref{obs}), 
and then exposures in the same filter are compared.
We arbitrarily assign one frame as a master, and compare each frame with the
master, calculating the magnitude differences for all the stars with flags of 
zero that the two frames have in common.
After adding the magnitude independent uncertainty derived later in this 
section, we take a weighted mean to find the relative transparency correction 
between this frame and the master.

Where there are $N$ frames, each frame is corrected by $N/(N+1)$ of its
relative transparency correction with respect to the master, and the master
is corrected by the sum of the transparency corrections divided by $N$.
Clearly, images taken in non-photometric conditions (such as the short
exposures in our wide survey) can be corrected to match the
photometric data in a similar way.
This process not only improves the internal consistency of the catalogue,
but results in an averaging of the profile corrections, improving the
absolute accuracy.
The transparency corrections derived should be small, and are a test of how 
accurate the profile correction is, and how photometric the conditions were.
The RMS of the transparency corrections applied for the deep survey is about
1 milli-mag, despite the fact that there was some variation in the extinction 
on the night on which they were taken.

\begin{figure}
\vspace{70mm}
\includegraphics{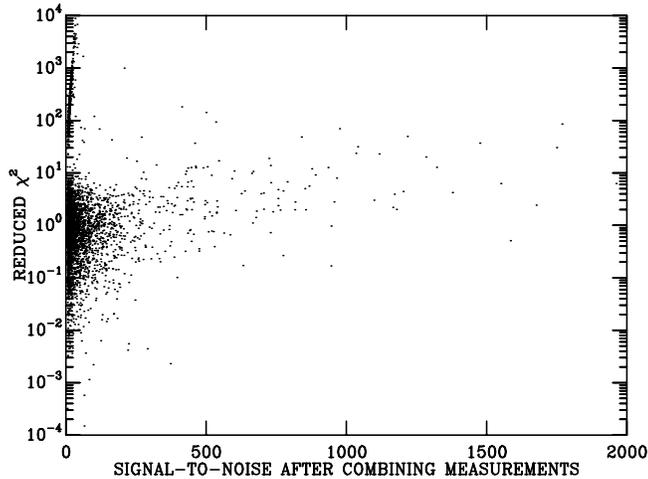}
\caption{
The $\chi^2 _\nu$ for each star in the three V-band exposures of deep field 1.
The number of degrees of freedom used was two.
}
\label{comb_stat}
\end{figure}

\begin{figure}
\vspace{70mm}
\includegraphics{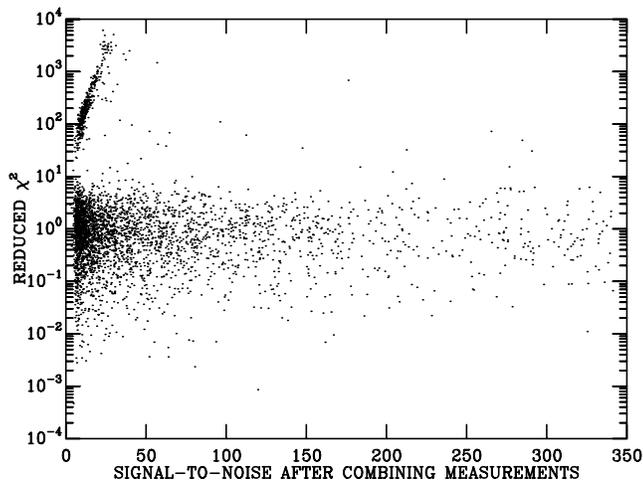}
\caption{
As Fig. \ref{comb_stat} but after the addition of a magnitude independent 
uncertainty of 5 milli-mags.
Note that this has the effect of reducing the calculated signal-to-noise ratio
for the brightest stars.
}
\label{comb_sys}
\end{figure}

We can now combine the repeated measurements of each star in a given filter
and at the same time check that our uncertainty estimates are correct.
The observations of an individual star are combined using a weighted mean
(in flux space), and at the same time a $\chi^2$ is calculated.
Fig. \ref{comb_stat} shows an example of these $\chi^2$s as a function of 
the signal-to-noise ratio in the combined measurement, ignoring the magnitude
independent uncertainty.
At signal-to-noise ratios of less than a hundred this plot matches the
expected distribution of $\chi^2$ very well, but there is a very clear
increase in $\chi^2$ at high signal-to-noise ratios, implying that we
are underestimating the uncertainty.
We can now estimate the magnitude independent uncertainty by adding an
increasing value in quadrature to the statistical uncertainty, until
the distribution in Fig.  \ref{comb_stat} becomes flat with
signal-to-noise ratio.
In practice the best way to achieve this is to plot histograms of the
distribution of $\chi^2$ above and below some signal-to-noise ratio
cut, and find the value of the magnitude independent uncertainty which
makes the two co-incide.
Fig. \ref{comb_sys} shows the same plot after the completion of this
process.  We can now create a final weighted mean and uncertainty for
each star, which includes the estimated magnitude independent
uncertainty for a single measurement.

The reader will no doubt be curious as to the origin of the data creating
the feature in the top left of Fig. \ref{comb_sys}.
These are due to weak cosmic ray events.
As they appear in just one of the images, they result in a high $\chi^2$ for
the star.
Stronger events have a higher $\chi^2$ and 
an apparently higher signal-to-noise ratio, hence the feature moves
from bottom left to top right.
It stops at the point at which the signal-to-noise ratio in the image is
sufficiently high that the procedure outlined in Section \ref{detect}
can be sure they are non stellar.
For obvious reasons this feature only appears in the $\chi^2$ plots for the 
waveband in which the detection algorithm was run.

It is clear that these spurious sources should have non-zero flags in the 
final catalogue.
The simplest way to achieve this is to flag all points with $\chi^2  _\nu$
above some critical value (we use $\chi^2 _\nu > 10 $ for the deep fields, 20 for the shallow
ones).
However, in addition to spurious sources, this will also pick out variable
stars, hence we refer to this as the variability flag.
As it is uncertain where a variable object should be placed on a 
colour-magnitude diagram, these stars, like all other objects with non-zero 
flags are not plotted on our standard CMDs, or included in our later
analysis.
(Despite the fact that pre-main sequence (PMS) stars are known to be variable,
only 3 objects flagged as variable lay close to our PMS.
This apparent lack of variability is probably because when we have repeat
images of the same field to the same depth, they are only minutes apart.)
It should be noted, however, that this is a very powerful way of detecting
variable stars.
It is more robust than the standard method of comparing the RMS scatter of a 
given object with ones of similar magnitude.
For example the RMS technique relies on the measurements at different epochs 
having similar uncertainties, whereas the $\chi^2$ method works with images of
very different effective exposure times.

\begin{figure}
\vspace{70mm}
\includegraphics{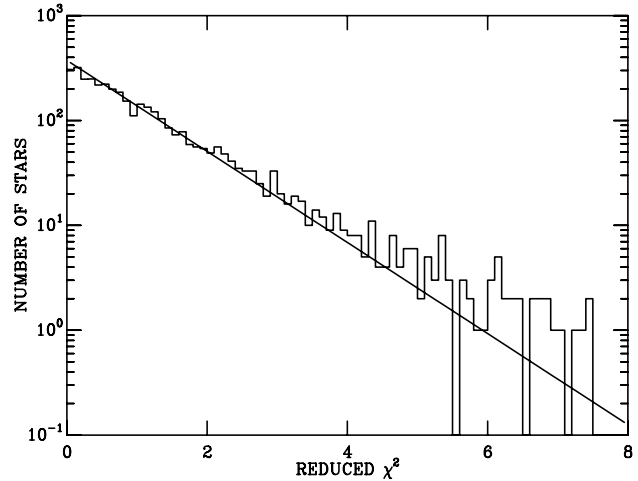}
\caption{
The histogram shows the distribution of $\chi^2$ for each star in the three 
I-band exposures of deep field 1.
The line is the expected distribution of $\chi^2$ for two degrees of freedom.
We emphasize that this is not a fit to the data (there are no free parameters).
}
\label{chi}
\end{figure}

We should emphasize that the reason we can can use $\chi^2$, is that
the optimal extraction provides very robust error estimates.
The data in Fig. \ref{comb_sys} allow us to test this, by 
comparing the distribution of $\chi^2$ for the stars, with that expected
if the uncertainties have been correctly calculated.
Fig. \ref{chi} shows that the measured $\chi^2$ distribution matches the
theoretical one well.
Even recognizing there is a small excess of sources with high $\chi^2$ 
(which could be genuine variability), the result is
clearly better than the only similar experiment we know of for profile
fitting photometry \citep[][Fig. 3]{caldwell91}.  

A few of our variable objects will be objects where an instrumental
problem, such as a ``cosmic ray'' has affected a single measurement.
With only three measurements we would be uncomfortable rejecting one 
discrepant point and presenting the mean of the other two as a measurement of
a varying object.
However, it is clear that given a few more frames, such a strategy may be 
useful.
In the case of the current dataset, we have to rely on the assumption that 
such events occur randomly across the field, and so the number of stars
lost to false variability will be accurately modeled by our completeness 
correction.

After the $\chi^2$ testing, the weighted mean instrumental magnitudes for
each filter were combined to give instrumental colours on a field-by-field
basis.
Since each instrumental colour originates from two mean instrumental 
magnitudes, it has two flags associated with it.
As these are one digit integers, the combined flag is defined as
ten times the blue flag plus the red flag.
The colours and magnitudes were then transformed onto 
the standard system using the photometric transformations whose derivation
is described in Section \ref{obs}.
Since the transformation for the $V$ magnitude involves a term in $B-V$, it
too acquires a two digit flag.

\section{ASTROMETRIC CALIBRATION}
\label{astrom}

The astrometry is performed using a model with six free parameters to transform
pixel co-ordinates into right ascension and declination.  
We use the SLALIB routines \citep{wallace99} to implement the method outlined 
in \cite{wallace98}.
This gives us the flexibility to include pin cushion/barrel distortion in
datasets where it is required, though a straightforward tangent plane
geometry was sufficient for the data presented here.
To fix the co-efficients of the model, we need a network of astrometric 
standards which provides at least tens of stars per CCD field.
None of the catalogues of primary astrometric standards provide such a network,
but there are several catalogues of secondary astrometric 
standards which have sufficiently dense coverage for our purposes.
We have used both the USNO-A2 \citep{monet98} and 2MASS catalogues for this 
work, but where it is available (as in the case of NGC2547), we prefer to use 
the SuperCOSMOS Sky Survey (SSS; 
\citeauthor{sss1} \citeyear{sss1};
\citeauthor{sss3} \citeyear{sss3}).
Its primary advantage over the other catalogues is the presence of
proper motion information, which we use to update the positions to the epoch 
of our CCD images.
We use the information as to whether stars are part of a blended image to
select isolated objects (Blend=0).
By default the catalogues supplied from the SSS do not include objects near 
bright stars.
In the case of NGC2547, such a restriction leads to large areas which contain 
no stars and so we have chosen to accept such objects by accepting
SuperCOSMOS quality flags less than or equal to 2047.

We find an initial astrometric solution for each CCD using the positions of 
three stars.
We then use this to pair SSS stars with objects in our fields, and then 
refine the solution using 3-$\sigma$ clipping.
In the case of NGC2547 the RMS residuals for each field were in the range
0.17-0.20 arcsec for a few hundred stars.

\section{FINAL PHOTOMETRIC CALIBRATION}
\label{fcal}

Once we have measurements of stars in the standard photometric system, with 
celestial co-ordinates, we can compare measurements of stars which occur
in two or more fields.
We calculated the mean magnitude differences in each overlap
region for such stars, and then adjusted the zero points for each
field to minimize these differences.
(Given the interlocking nature of the grid of fields these differences can 
only be minimized, not entirely removed.)
Not only does this process improve the consistency between fields in the final 
catalogue, but the RMS which remains after this process gives us an
estimate of systematic errors with position in the profile correction.
In this process we ensured that the mean shift of the nine wide survey fields
was zero, since their photometric calibration was tied to their deep
exposures, which were taken under photometric conditions.
After the adjustment process, the mean differences for the 42
overlaps in the combined dataset were $V$=0.014, $B-V$=0.005 and $V-I_c$=0.011.
%The adjustment procedure should eliminate errors in the profile
%correction between fields, such as choosing a large aperture too small
%to take account of seeing changes.

The uncertainties which appear in the catalogue are those calculated at the
end of Section \ref{comb_meas}.
It is important that we are clear that this uncertainty represents the scatter 
between measurements that would be expected were the same field observed 
repeatedly.
They do not include any systematic position-dependent errors in our profile 
correction.
For example, if we used too few terms to represent the profile correction as
a function of position, then (assuming the seeing remained the same) it would 
be systematically wrong in the same sense in every frame.
Thus the uncertainty in the catalogue should be used to compare the
magnitudes of stars at similar positions in the field.
If one wishes to compare stars at, say, opposite sides of the field,
or from different fields, then one must add the error derived from the 
overlap between fields.
Finally, if one wishes to compare our photometry with other work, the
uncertainty in our photometric calibration should be added.

\section{THE FINAL CATALOGUES}
\label{cats}

\begin{figure}
\vspace{70mm}
\includegraphics{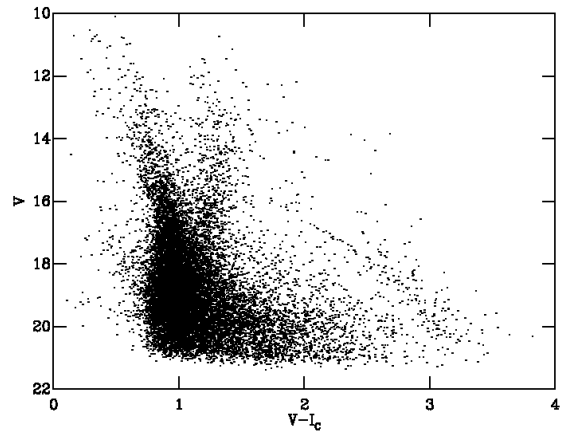}
\caption{
$V$ vs. $V-I_c$ for the combined catalogue.
Sources with a signal-to-noise ratio worse than 10
have been omitted.
}
\label{comb_vi}
\end{figure}

\begin{figure}
\vspace{70mm}
\includegraphics{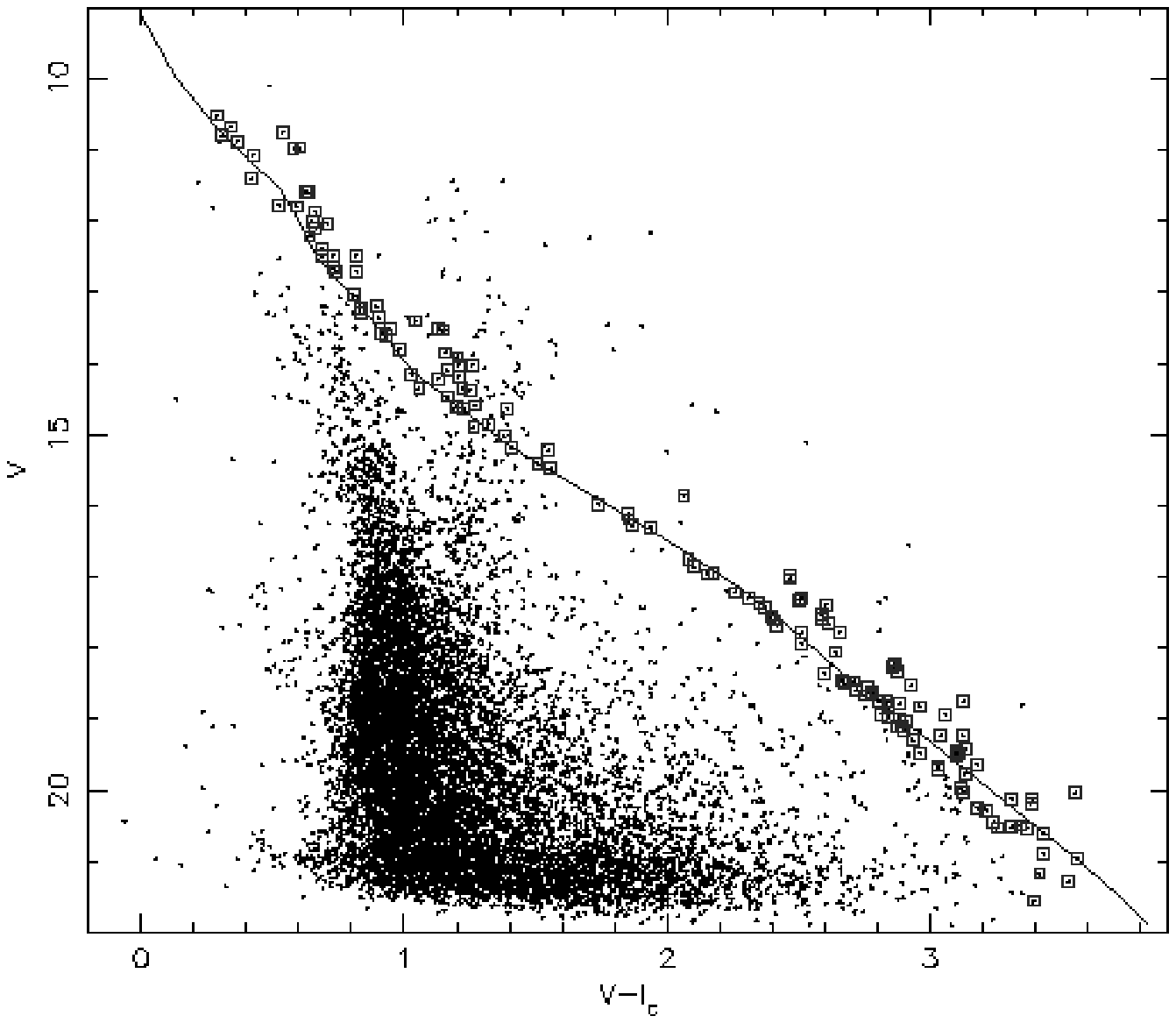}
\caption{
$V$ vs. $V-I_c$ for the enhanced deep catalogue. The solid line is the
best fitting D'Antona \& Mazzitelli isochrone discussed in
Section~\ref{age} and the points enclosed by squares are candidate
cluster members selected in Section~\ref{cands}.
Sources with a signal-to-noise ratio worse than 5
have been omitted.
}
\label{deep_vi}
\end{figure}

\begin{figure}
\vspace{70mm}
\includegraphics{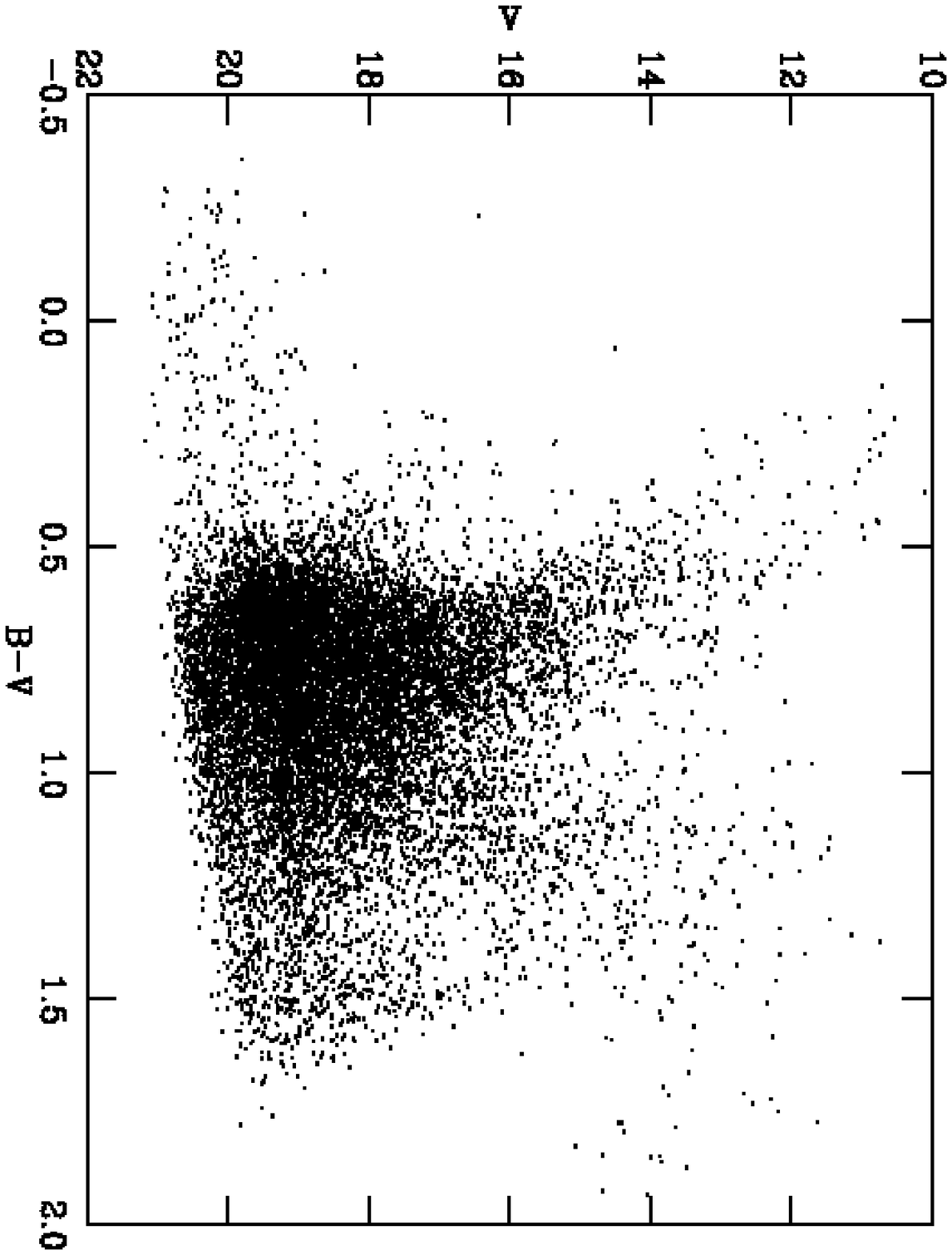}
\caption{
$V$ vs. $B-V$ for the combined catalogue.
Sources with a signal-to-noise ratio worse than 10
have been omitted.
}
\label{comb_bv}
\end{figure}

Given the different depths of the two surveys, it was found useful to 
keep three separate catalogues, one for the four deep fields, one for
the nine fields of the wide survey, and one which 
is the result of  combining all 13 fields.
We will refer to these as the deep, wide and combined catalogues respectively.
(Note that the corrections derived from the overlaps and applied to each field
are, in all cases, those derived from combining all the fields.)
The wide and combined catalogues cover approximately 0.32 square degree, the
deep catalogue 0.12 square degree.
In Figs. \ref{comb_vi} to \ref{comb_bv} we show CMDs from these catalogues, 
restricted to stars which have flags of zero in each co-ordinate.
In Table 2 we present an example catalogue, which is the stars in common 
between the combined catalogue and \cite{claria82}.
We choose this as our example as it allows easy cross identification between
our own work and that of Clari\'a, which would otherwise be difficult as
there are only finding charts, not positions in \cite{claria82}.

\begin{figure}
\vspace{70mm}
\includegraphics{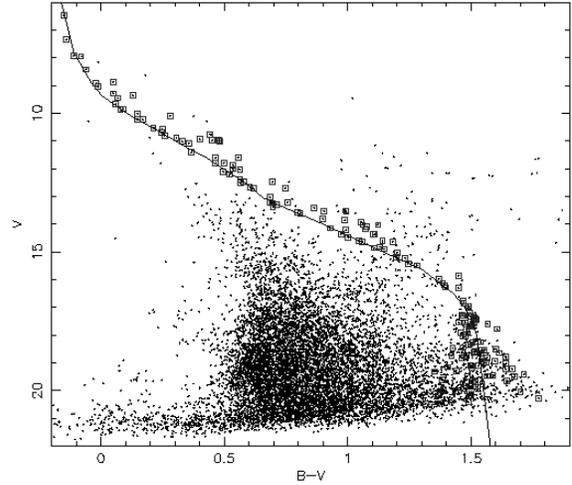}
\caption{
$V$ vs. $B-V$ for the enhanced deep catalogue. The solid line is the
best fitting D'Antona \& Mazzitelli isochrone discussed in
Section~\ref{age} and the points enclosed by squares are candidate
cluster members selected in Section~\ref{cands}.
Sources with a signal-to-noise ratio worse than 5
have been omitted.
}
\label{deep_bv}
\end{figure} 

In magnitude space, the error bars for an object are two sided.
We have chosen to quote those for the bright side of the data point for the
following reason.
It is desirable that the bright side error bar for objects with positive 
fluxes matches
smoothly onto the upper limit for objects with zero flux.
For this to be true it means that as the flux for an object tends to zero,
the result of subtracting the magnitude uncertainty from the magnitude (i.e. 
calculating the bright end of the error bar) must tend towards the value 
obtained by converting the flux uncertainty into a magnitude.
This forces our choice of the bright-side error bar.
However, for situations where the faint side error bar is important, it is
trivial to show that it can be calculated from the uncertainty we give ($E$) as
\begin{equation} 
-2.5log_{10} (1 - (10^{E/2.5} -1)^2). 
\end{equation}

Since the deep survey did not contain data for the brighter stars, we obtained
this from both the wide survey and the data of \cite{claria82} in the 
following way.
We first extracted the saturated stars from the deep catalogue, and then 
searched at these positions for stars in the wide survey.
Where magnitudes with flags of zero were found, these were inserted into the 
deep catalogue, and the stars deleted from our list of saturated objects.
We then searched Clari\'a's catalogue for counterparts for the remaining 
objects in our saturated list, and placed these into the deep catalogue as 
well.
We refer to the result of this process as the enhanced deep catalogue.
We resisted the temptation to replace entries with the bad pixel flag
in the deep catalogue with entries from other catalogues, as this would have
made the area of sky surveyed a function of magnitude.

In a similar way we replaced all saturated objects in the wide and combined 
catalogues
with magnitudes from \cite{claria82}, to create an enhanced combined catalogue
and an enhanced wide catalogue.
It should be remembered that \cite{claria82} only contains data in $V$ and 
$B-V$, and thus the enhanced catalogues reach brighter magnitudes in $V$ and 
$B-V$ than $V-I_c$.
The enhanced deep, wide and combined catalogues have been deposited with 
Centre de Donn\'ees astronomiques de Strasbourg as Tables 3, 4 and 5 
respectively.
They contain all the columns of Table 2, except for the Clari\'a identification
numbers. 

One striking feature of the $V$,$B-V$ CMDs is the appearance of faint blue 
objects.
Since similar objects also appear in the wide catalogue, but at slightly
brighter magnitudes, they are clearly spurious.
Indeed, comparison of Figs. \ref{comb_bv} and \ref{deep_bv}, show how there
are many more such objects at lower signal-to-noise ratios.
Close examination shows that these spurious objects are often associated
with bright stars, which are so faint that they cannot reliably be classified
as non-stellar.
The fact they appear blue is presumably related to the nature of the 
scattering in the optics which produces them.
These objects produce a sharp increase in the number of objects which appear
in the wide catalogue, but not in the deep catalogue (and are therefore
spurious), below $V$=20.0, or an uncertainty of 0.08 mags.
Approximately 30 percent of objects below this limit are spurious.
However, as the corresponding point in signal-to-noise ratio in the deep catalogue 
is V=20.7, this can only affect the last bin of our mass function
derived in Section \ref{func}, and the largest possible effect is less than 
the error bar. 

\subsection{Comparison with previous work}

\begin{figure}
\vspace{70mm}
\includegraphics{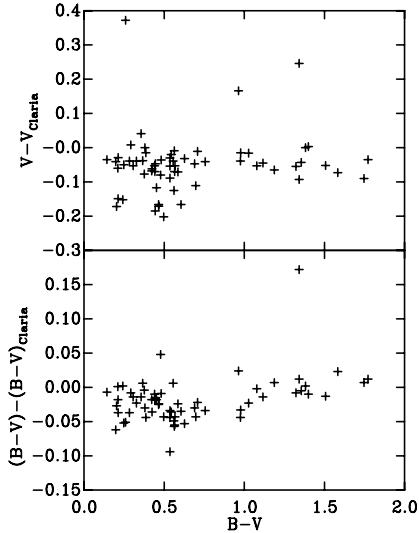}
\caption{The difference between the $V$ magnitudes and $B-V$ colours we
derived and those of Clari\'a, as a function of our $B-V$.
Although the $B-V$ differences appear to be a function of colour, a linear
term is only significant at the 73\% level.
}
\label{comp_claria}
\end{figure}

The best comparison we can undertake is with the photo-electric photometry
of \cite{claria82} since, in the magnitude range in question,
the uncertainties in both our own and Clari\'a's photometry
are independent of magnitude.
Of 118 stars in \cite{claria82} between $V$=5.6 and $V$=13.98, 102 lie within 
the area of the combined catalogue, and all are detected.
Of these objects a total of 39 were some combination of saturated, 
lying on bad pixels or non-stellar.
The remaining objects are plotted in Fig. \ref{comp_claria}.
Three have large differences between the two catalogues which can be 
attributed to nearby companions in each case.
This leaves 60 objects for comparison.
In $B-V$ we find a RMS difference of 0.024 mags, and an (unweighted) mean 
difference $(B-V) - (B-V)_{Claria}=-0.021$ mags.
The RMS difference is acceptable as the uncertainty in an individual star's
magnitude is dominated by the profile correction (0.005 mags, Section
\ref{fcal}, though in some cases the statistical uncertainty reaches 0.02 mags)
in our data, and is about 0.02 mags in the photoelectric work \citep{claria82}.
The mean difference is acceptable as the combination of the uncertainties in 
the photometric calibration of the two datasets.

There appears to be a problem with the V comparison, in that eight of
the remaining 60 objects form a group which appear about 0.1
magnitudes brighter in our data than Clari\'a's.
Although seven of the eight stars are on our field 18, this is clearly not
a problem with that field, as two of the eight stars appear on other fields,
and the differences between the two measurements are two hundredths or less,
and within the error bars.
Furthermore, if we examine the region of the PMS where it is best defined,
around $V-I_c =2$, shifting field 18 by 0.1 magnitudes makes the PMS members in
that field lie above those in the rest of the catalogue.
We conclude that the \cite{claria82} measurements are incorrect, but do so 
reluctantly as there is not an obvious reason why this should be so.
Removing these outliers we find an RMS difference of 0.032 mags and an
average difference of $V - V_{Claria}= -0.047$ mags.
Again the RMS is acceptable as the sum of our profile correction of
0.014 mags, and \cite{claria82}'s estimated uncertainty of 0.02 mags.
The mean difference in $V$ is disturbing, but is
similar in magnitude (though opposite in sign) to the difference between 
photoelectric photometry and modern CCD photometry found by \cite{hippel98} 
for NGC188.
%CL114 has a companion about 3 arcseconds way 0.15 mags fainter.
%CL104 companion 16 arcsec away 0.5 mags fainter.

It is also important to note that there is an almost one-to-one correspondence
between saturated objects in our catalogues and objects in the 
\cite{claria82} photometry.
This is because our detection algorithm does not break single saturated
stars into multiple objects.
The only problem is that the derived position for a few very badly
saturated objects (some are 10,000 times above our saturation limit) is 
sometimes much worse than the 1.5 arcsec correlation radius we use for 
identifying the same object observed in two or more fields.
Fortunately such objects are easy to identify, as they are very bright stars
at similar positions but in two separate fields.
This one-to-one correspondence is crucial, since it means we can be confident 
we have a complete count of all objects, even though they may be saturated in 
our data.

\begin{figure}
\vspace{70mm}
\includegraphics{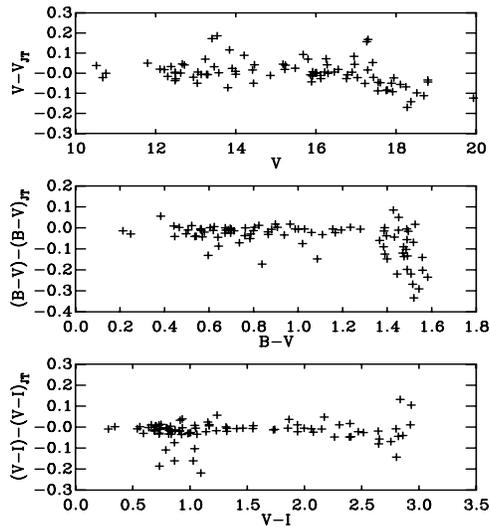}
\caption{The difference between the $V$ magnitudes and $B-V$ and $V-I$ 
colours we derived and those of Jeffries \& Tolley.
Between two and four data points lie outside the Y-axis scaling of each panel.
}
\label{comp_jt}
\end{figure}

Fig. \ref{comp_jt} shows the comparison between 
our combined catalogue and the results of \cite{jt98}.  
The interest of this comparison is that \cite{jt98} used the same
data, but reduced by more conventional (aperture photometry) algorithms.
Here, we reach sufficiently faint magnitudes that the error bars are
significantly different for some stars, and thus use weighted mean differences.
In addition we remove some outliers.
We find a mean difference of $(V-I_c) - (V-I_c)_{JT}=-0.009$.
The behaviour for $V$ and $B-V$ is more complex.  
$V-V_{JT}=-0.011$ for $V<$17, 
but thereafter $V-V_{JT}$ falls linearly, reaching about -0.15 at V=19.
Similarly, $(B-V) - (B-V)_{JT}=-0.012$ for $B-V<$1.33, but 
then falls to -0.25 by $B-V$=1.8
The \cite{jt98} stars are mainly along the pre-main-sequence, and thus $V$ is 
strongly correlated with $B-V$, so it is difficult to be certain whether the
above correlations are with $V$ or $B-V$, however, they are tightest in the
forms stated above.

Perhaps most the important comparison is that of the signal-to-noise ratio 
achieved in our reduction, and that of \cite{jt98}.
For a $V$=19 star which appears in just one of the longer exposures of the 
wide survey \cite{jt98} give an uncertainty of about 0.06 magnitudes
compared with our uncertainty of 0.04 magnitudes; an improvement of 1.5 in 
signal-to-noise ratio. 
This means that at a given signal-to-noise ratio, our CMDs reach about 0.45 
magnitudes deeper than the aperture photometry. 
%The aperture used in \cite{jt98} was 6 arcseconds diameter.
%Given that the mean FWHM seeing in the wide survey's long exposures was 1.5 
%arcseconds, this corresponds to an sperture of 2 FWHM radius.
%From examination of Fig. 1 of \cite{p1}, we expect an optimal extraction to 
%deliver about twice the signal-to-noise of such an aperture.
%In fact we have only obtained about a factor of 1.5 improvement

\section{X-RAY SELECTION, AND THE AGE OF THE CLUSTER}
\label{age}

\begin{figure}
\vspace{70mm}
\includegraphics{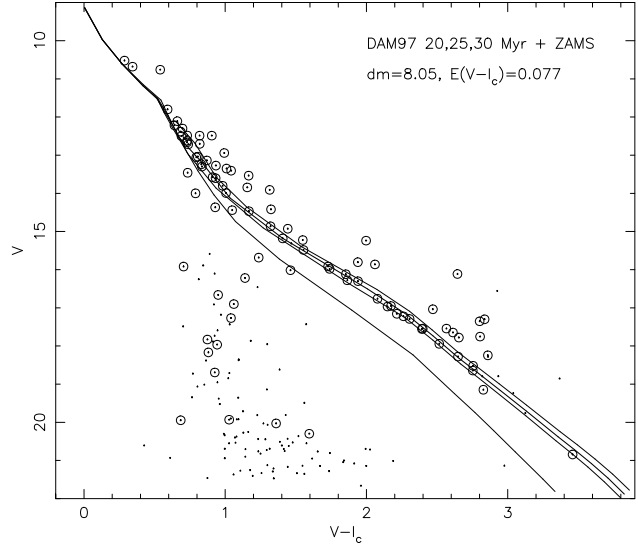}
\caption{
The CMD for the stars in the enhanced combined catalogue within X-ray error 
circles. 
Every star in an X-ray error circle is marked with a dot, with the dot 
for the brightest object within an error circle, circled. The solid lines
are isochrones derived from the models of D'Antona \& Mazzitelli (1997).
}
\label{xray}
\end{figure}

\begin{figure}
\vspace{70mm}
\includegraphics{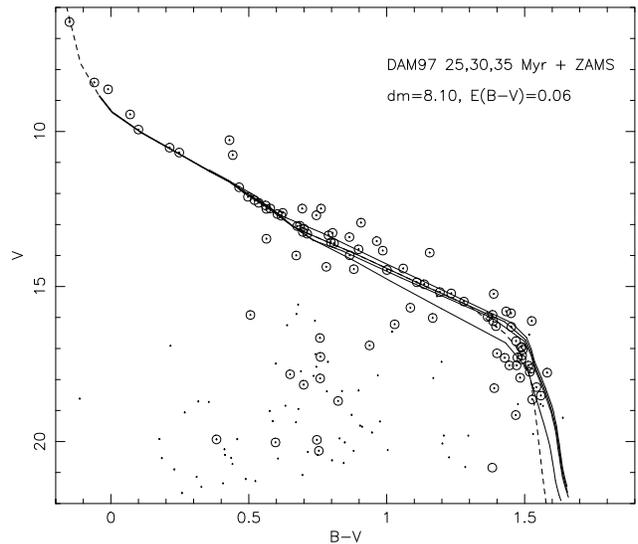}
\caption{
As for Fig. \ref{xray}, but in $V$ vs $B-V$.
The brighter members do not appear
in Fig. \ref{xray} as they lack $I$-band photometry.
The solid lines
are isochrones derived from the models of D'Antona \& Mazzitelli
(1997). The dashed lines show the extensions and alterations to the
30\,Myr isochrone that have been used to select candidate members in
Section \ref{cands} (see Section \ref{age}).
}
\label{xrayB}
\end{figure}

To determine the age of the cluster we fitted an isochrone to the X-ray 
selected members.
This is essentially an updated version of the analysis of \cite{jt98}, 
though with a deeper catalogue.
In addition, by repeating the analysis, we ensure that the mass function 
we will derive in Section \ref{func} is 
based on an isochrone which is consistent with the new reduction of the 
photometry.
We therefore cross-correlated our enhanced combined catalogue against the 
X-ray 
catalogue of \cite{jt98}, using the error radius appropriate to each 
X-ray position. 
We extracted all the stars within each error circle, and noted which was
the brightest within each circle.
The resulting CMDs are shown in Figs. \ref{xray} and \ref{xrayB}.

There are a small number of sources, which lie close to the PMS, which are
not the brightest objects in their error circles.
As this could be either a sign that the PMS stars are spatially clustered, 
or that a brighter non-member appears in the error circles by chance,
we examined the five such objects fainter than V=18.5 close to or above
the PMS.
Three objects have brighter stars in their error circles which are clearly 
not members, whilst two have brighter PMS objects in their error circles.
To establish if the 
latter are consistent with a random distribution of sources on the sky,
we tried shifting the error circles by +/-30 and 60 arcseconds in declination.
In these randomly placed error circles, we found an average total of 0.5 sources
in each of our four simulated datasets at positions in the CMD consistent 
with the PMS.
Although 0.5 is rather different from two, they are consistent within the
uncertainties, establishing that there is no clustering of the PMS stars.

With this issue clarified, we then began fitting the X-ray selected
members.  The methods that we use to generate model isochrones to fit
the colour-magnitude diagrams are described in great detail by
\cite{jth01}.  Briefly, we generate a set of model isochrones based
upon the solar metallicity models of \cite{dantona97} and
\cite{siess00}.  These models cover {\em most} of the mass range
required to interpret our NGC 2547 data, although the D'Antona \&
Mazzitelli models are limited to 3$M_{\odot}$ at the high mass end.
These isochrones are converted into the observational colour-magnitude
plane, using empirical $T_{\rm eff}$ to colour conversions, derived by
assuming that the Pleiades defines an isochrone at an age of 120\,Myr
and distance of 130\,pc \citep{stauffer98}. Note that if a different
fiducial distance were adopted for the Pleiades, this would change our
derived distance for NGC 2547 by a similar percentage.  In addition we need
to assume relationships between bolometric correction and colour, which
come from \cite{flower96} for $B-V$, \cite{leggett96} for $V-I_{\rm
c}>0.7$ and the atmospheric models of \cite{bessell98} for $V-I_{\rm
c}<0.7$.

These isochrones are then compared to the objects which correlate with
X-ray sources. As discussed by \cite{jt98}, these are
very likely to be cluster members with very little contamination by
non-members, so should allow us to determine the cluster age, distance and
reddening. For the purposes of this paper we choose to fix the
reddening at the value of $E(B-V)=0.06$, determined from the hot stars
by \cite{claria82}. 
We further assume a standard reddening law and that
$E(V-I_{\rm c})=0.077$. Having done this, we find that the age and
distance are only partially degenerate parameters. The distance is
reasonably well determined from the hotter ZAMS stars, especially in
the $V$ vs $B-V$ diagram, whereas the age is
determined by the positions of the cooler stars which are still
descending PMS tracks towards the ZAMS. Any effort to find the best-fit
age and distance, by minimising chi-squared for example, is fraught
with difficulty, because many of the X-ray sources appear to be either
non-members, or are binary members of the cluster and lie above any
single star isochrone. However, by eye, there does appear to be a clear
sequence of about 30 stars in each CMD which we assume defines the
cluster single star isochrone. We proceed by choosing isochrones of
different ages and then adjusting the distance modulus to get what
appears to be the best possible fit to this sequence. The uncertainties
of this process are mostly due to the degeneracy between distance and
age, but because the isochrones change {\em shape} with age, we find
that beyond certain limits we cannot adjust the distance and get a reasonable
match between the data and the model.

\begin{figure}
\vspace{70mm}
\includegraphics{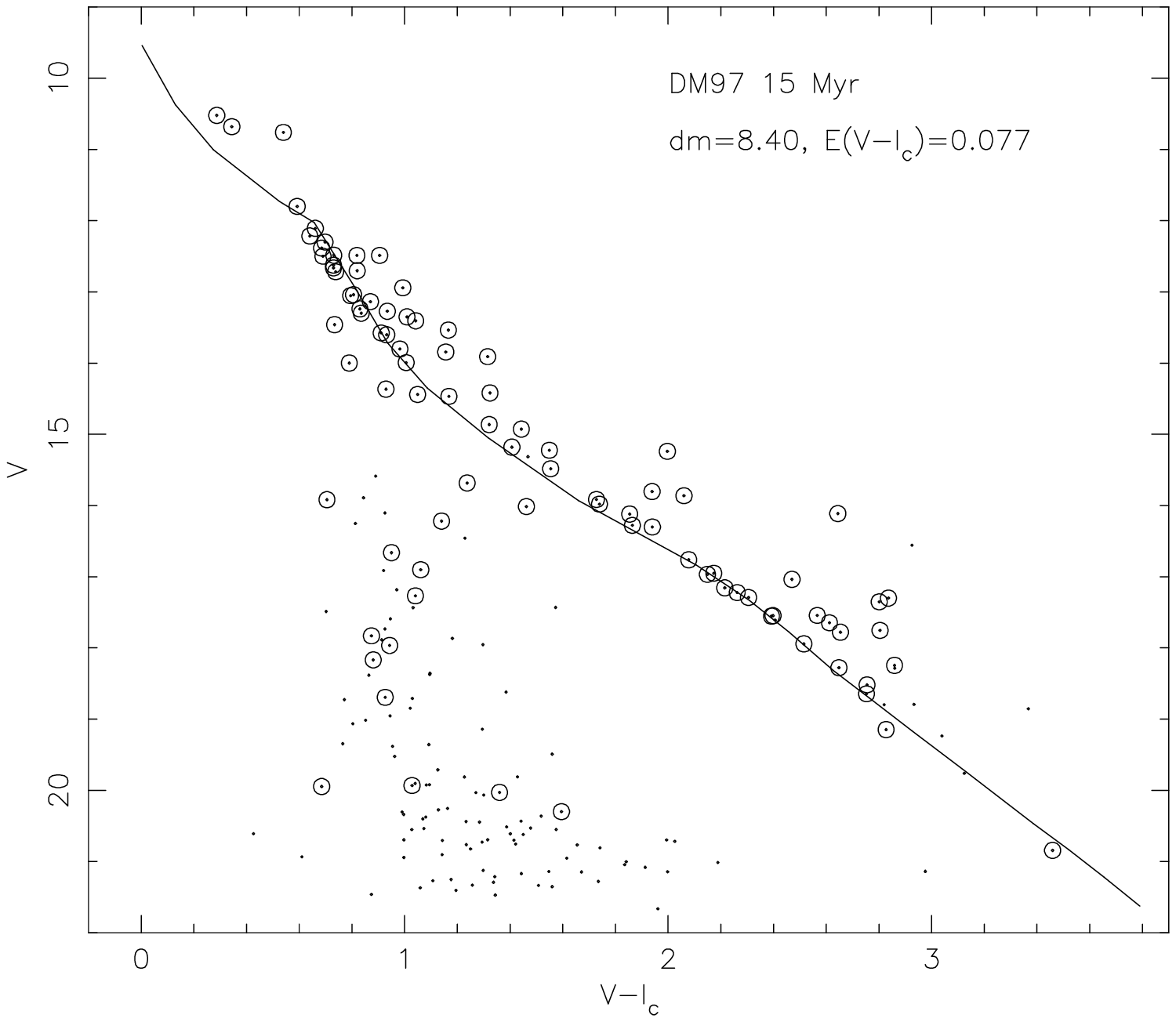}
\caption{
As for Fig. \ref{xray}, but showing the poor fit given by an age of 15Myr with
the models of D'Antona \& Mazzitelli (1997).
}
\label{xray2}
\end{figure}

\begin{figure}
\vspace{70mm}
\includegraphics{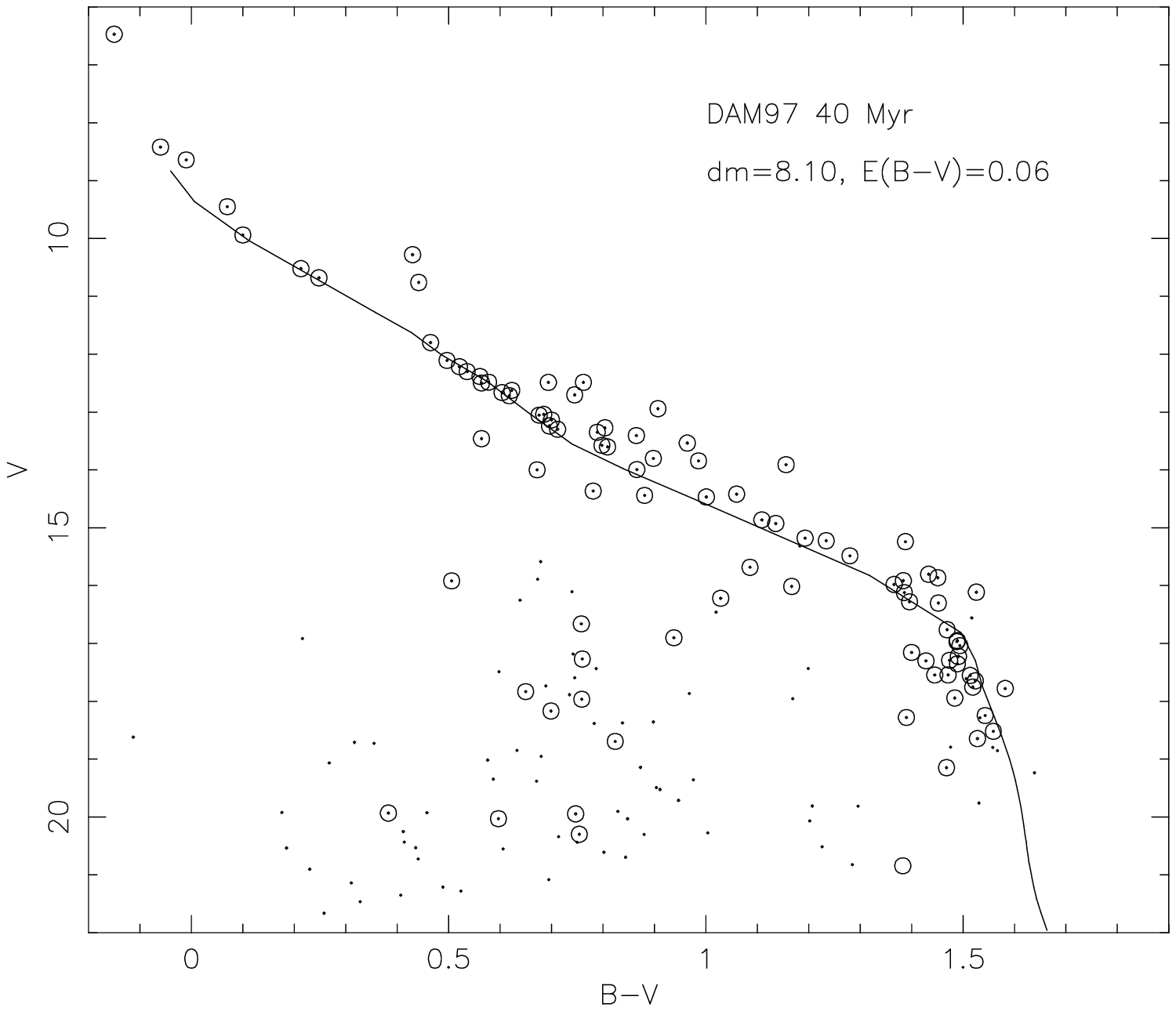}
\caption{
As for Fig. \ref{xrayB}, but showing the poor fit given by an age of 40Myr with
the models of D'Antona \& Mazzitelli (1997).
}
\label{xrayB2}
\end{figure}

The best fitting \cite{dantona97} models are shown in Figs. \ref{xray}
and \ref{xrayB}.  For $V$ vs $B-V$ diagram we find an intrinsic
distance modulus of $8.10\pm0.05$ and an age of $30\pm5$\,Myr. For the
$V$ vs $V-I_{\rm c}$ diagram we find an intrinsic distance modulus of
$8.05\pm0.10$ and an age of $25\pm5$\,Myr.  The Siess et al. (2000)
isochrones are similar but we find ages approximately 5\,Myr
older. That the distance modulus in the $V$ vs $B-V$ CMD may be
slightly larger than for the $V$ vs $V-I_{\rm c}$ CMD, might be
attributable to a slightly subsolar ([Fe/H]$\sim-0.1$) metallicity for
NGC 2547 \citep[see][for discussion]{jth01}. If this were the case the
intrinsic distance modulus would be lowered to 7.95-8.0.  Either range
of distances is consistent with that derived from the Hipparcos data of
$8.18^{+0.29}_{-0.26}$ \citep{robichon99}.  To illustrate the validity
of our uncertainty estimates, Figs.~\ref{xray2} and \ref{xrayB2} show
our best efforts to obtain fits at both younger (15\,Myr) and older
(40\,Myr) ages, using the \cite{dantona97} models. The younger age is
ruled out using the $V$ vs $V-I_{\rm c}$ CMD. The shape does not really
match the observed sequence and the required distance modulus is {\em
much} larger than the $8.10\pm0.05$ defined by the hotter ZAMS stars in
the $V$ vs $B-V$ CMD. The older age is ruled out by the $V$ vs $B-V$
CMD. The stars between $0.8<B-V<1.4$ cannot be simultaneously fit with
the higher mass objects that have reached the ZAMS.
Thus we conclude the age lies between 20 and 35Myr, using the \cite{dantona97} 
models.

The value for the distance is similar to that quoted by \cite{jt98},
but the age is somewhat older (c.f. $14\pm4$\,Myr), even though the
same \cite{dantona97} evolutionary models are used in that paper. We
attribute this difference to the use of a relationship between
bolometric correction and $T_{\rm eff}$ in \cite{jt98}, to convert
luminosities to absolute $V$ magnitudes -- a practice commonly found in
the literature.  The method described above and in \cite{jth01} uses
the {\em much} more reliable conversion between bolometric correction
and {\em colour}.

We note that the fit of the isochrones in the $V$ vs $B-V$ CMD is poor
around $B-V\simeq1.5$. This may be a problem with the evolutionary
models or a problem in the relative calibration of our photometry with
that of the Pleiades used in defining the $B-V$ effective temperature
relation. We note that we have plenty of well measured standards beyond
$B-V=1.5$ and that our transformation from instrumental to standard
$B-V$ is therefore well constrained. 
In order to define an isochrone
that can be used solely to select cluster candidates we have arbitrarily defined an
isochrone that departs from the 30\,Myr D'Antona \& Mazzitelli
isochrone at $B-V>1.35$ and which passes through the bulk of the X-ray
sources (see Fig.11). In the same spirit, we have also extended the
model isochrone to higher masses.

\section{Selection of cluster candidates}
\label{cands}

\subsection{Candidate selection}

Selection of candidate members of NGC 2547 proceeds exactly in the way
described in detail by \cite{jth01}. Each star in the catalogue is tested
against a number of criteria for membership. If a star does not fail
any of these tests then it is retained as a candidate member. Briefly,
these tests consist of investigating whether a star is: (a) close to the
$V$ vs $V-I_{\rm c}$ isochrone defined in Sect.11; (b) close to the $V$
vs $B-V$ isochrone; (c) close to the $V-I_{\rm c}$ vs $B-V$ locus for
cluster members. We reject those measurements which have a non-zero
flags or uncertainties of greater than 0.2 mag. A star must have at least a
$V$ magnitude and one measured colour to be classed as a candidate
member. In practice, there are some differences in the membership list
depending on whether we use the \cite{dantona97} or the
Siess et al. (2000) isochrones. Despite having distance and age as free
parameters, the two sets of evolutionary models give isochrones that
have slightly different {\em shapes}, particularly for the very cool stars.

In addition we apply a final test for possible binarity. If a star lies
more than 0.3 mag. above the $V$ vs $V-I_{\rm c}$ isochrone (or the $V$
vs $B-V$ isochrone for stars with $V-I_{\rm c}<0.5$) we class them as
candidate unresolved binary systems. To calculate what mass ratios we
are sensitive to, we have used the slopes of our CMDs and
mass-magnitude relationships (see Section~\ref{func}), and equation~5
from \citep{kahler99}. We find that $\Delta V\geq0.3$ corresponds to a
mass ratio $q\geq 0.6$ for $V<14$ and $q\geq0.5$ for fainter cluster
members. The reason for the difference is the change in slope of the
CMDs and mass-magnitude relation as PMS stars reach the ZAMS. In
contrast, for older clusters like the Pleiades, $\Delta V\geq0.3$
corresponds to $q\geq0.6$ over the entire mass range considered here.

The reader might suspect that by choosing only 
objects flagged as stellar, we may have excluded wide binaries that would appear as
non-stellar objects in the catalogue. In fact, given the typical seeing
of our data and the distance of the cluster, equal mass binaries would
have to be separated by more than a few tenths of an arcsecond (more
than 100\,AU) and unequal mass binaries by even more. 
\cite{bouvier97bin} estimate that the fraction of such systems among the
intermediate and low-mass stars of the Pleiades is at most a few
percent, so it is possible we have rejected these systems as well as
triple (or even higher multiple) systems that might fall outside our
selection criteria in the colour-magnitude diagrams \citep[estimated as less
than 2 percent of systems -- ][]{mermilliod92}. Incompleteness of
this order will have no great effect on any of our results or conclusions.

We used the D'Antona \& Mazzitelli isochrones to find 184 cluster
candidates from our enhanced deep catalogue, of which 52 are probable
unresolved, high mass ratio binaries.  The Siess et al. (2000)
isochrones yield 201 cluster candidates, of which 61 are candidate
binaries. The additional members in the Siess et al. case are mainly at
faint magnitudes (see below).  
Figs. \ref{deep_vi} and \ref{deep_bv} show the filtered, deep $V$ vs 
$B-V$ and $V$ vs $V-I_{\rm c}$ CMDs, with both the loci used to select 
the cluster candidates and the selected cluster candidates indicated
(for the D'Antona \& Mazzitelli isochrones).
Tables presenting ID and field numbers (data from \cite{claria82} are given 
as field zero), equatorial co-ordinates, membership and binary flags
have been deposited with Centre de Donn\'ees astronomiques de Strasbourg 
as Tables 6 and 7 for \cite{dantona97} and \cite{siess00} models
respectively.

\subsection{Field star contamination}

We are confident that our membership selection criteria have included
almost all the true cluster members. However, despite having a cleanly
defined sequence, especially in the $V$ vs $V-I_{\rm c}$
colour-magnitude diagram, it is inevitable that we have included
contaminating field stars as candidate members. \cite{jth01} devised a scheme
to estimate the size of this contamination (as a function of colour),
by interpolating the distribution of background stars over the interval
from which cluster candidates were selected. Unfortunately in NGC 2547,
there are far fewer stars immediately above and below the cluster
sequence with which to do this interpolation, and we find that this
technique will not work, apart from in the region where background
giants intrude at $14.0<V<15.5$. The total number of stars in the NGC
2547 catalogues is comparable to those discussed by \cite{jth01} for NGC 2516,
an older cluster ($\sim 100-150$\,Myr) at a similar distance. It seems
then that the numbers of contaminating field stars in the NGC 2547 must
be much lower than for NGC 2516, despite NGC 2547 being at a lower
galactic latitude ($b=-9^{\circ}$ and $-16^{\circ}$ respectively). We
believe this is (a) because we have observed a smaller area in NGC 2547
and (b) because NGC 2547 is younger than NGC 2516 and so, at a similar
distance, the NGC 2547 isochrone for cool PMS stars lies more than a
magnitude above the ZAMS in the colour-magnitude diagram and hence
clear of the background contamination.

For the rest of this paper we will make the assumption that
contamination of our NGC 2547 cluster sample is negligible. The
exception is between $14.0<V<15.5$ where integration of an
interpolation of the density of stars above and below the cluster
sequence, suggests that there, the field star contamination could be as
high as 40 percent. Note that this assumption was recently tested for
the brighter stars ($12<V<14$) by \cite{jtj00}.
They performed high resolution spectroscopy of 23 of the stars
selected as cluster candidates in this paper. From the radial
velocities there is good evidence that at least 20, and probably all, of
these stars are cluster members.

\section{COMPLETENESS -- GHOST STARS}
\label{ghosts}

Before we can construct a luminosity function from our candidate list we
must determine the fraction of stars
at a given magnitude which we detect and for which we successfully obtain 
photometry. 
Examination of the reduction path shows the following hurdles a PMS star
must overcome to appear in our final catalogue.
First it must be detected in our V-band images.
Most simply, it may be too faint, but as 
the point-spread function is a function of position, stars of a given
magnitude are less likely to be detected at the frame edges than at its
center.
It may also not be detected because it
lies close to a brighter star, or a chip defect.
Next, our test for point-like objects (Section \ref{detect}) will reject
some small proportion of star-like objects.
The detected object will next pass through the photometry routines, when, to
appear in the final catalogue, it must avoid being flagged for saturation or
proximity to chip defects or a CCD edge.
Very occasionally a star will also acquire a non-zero flag because the fit to 
the sky histogram fails. 
When several measurements of the same object are combined the
star may be rejected on the grounds of variability. 
Finally, a PMS star which appears in our final catalogue, may be rejected
because it falls outside our selection strip.

The most convincing way to account for this mass of selection criteria
is to inject a simulated sequence of PMS stars into the data and examine 
what fraction appear in our final catalogue, and are selected as PMS stars.
This has the advantage that we simulate our completeness as a function of 
colour, as well as magnitude.
(This is important, not only because of colour terms in the photometric
transformation, but also because the error bars in the colours are a function
of colour as well as magnitude.)
As we have not modeled the PSF as a function of position, we cannot 
generate stars at random positions on the CCD. 
Instead we use the following procedure, which we term ``ghosting''.
The essence of this technique is that images of the brightest stars in the 
field are sky-subtracted, scaled to the required brightness for a simulated 
star, and copied to a nearby position.
The position has to be sufficiently close that the PSF has not varied 
significantly, but far enough away to avoid the wings of the original.

\subsection{The area to be copied}

The most significant issue with ghosting is deciding how large an area 
must be copied.
The detection technique relies on the counts within a sliding box being
greater than the detection threshold.
As long as the total counts in the box lie above the threshold when the
box is at the peak of the star, the star will be detected.
The wings of the star are irrelevant.
Thus for the detection algorithm to work, an area equal to the detection
box (typically one FWHM) must be copied.
The position of the star is determined by fitting it \citep[see][]{p1}, and
the box used for this is twice the FWHM (corresponding to a ``fitting radius''
of a FWHM), so again only a small area needs to be copied. 
For the photometry to be successful, the copied area must cover the weight
mask out to the radius at which it is clipped, normally two FWHM.
Thus it is the requirements of the photometry which decide the area to be
copied, and which normally sets it to two FWHM.
If we copy the star out to two FHWM radius, can we be sure that
the original is point-like at this radius?
The answer must be yes, since otherwise it would have been flagged as 
non-stellar.
Despite the above reasoning, the brighter ghosts can be quite striking in
appearance, with a discontinuity at the edge of the copied area.
However, this does not affect their detectability, or measured brightness.

\subsection{Practicalities of the simulation}

To create the ghosts, we first create a list of the magnitudes and colours 
of the stars we wish to simulate.
We then select a parent to be used for each ghost, scale it appropriately,
and insert it into each image at a given offset from the parent.
The most obvious way to decide the scaling is to use the measured photometric
transformations to calculate the number of counts required in the ghost.
However, a whole series of minor corrections, such as the relative transparency
correction (Section \ref{comb_meas}), have been applied to the data.
It is simpler, therefore (and hence more robust against coding errors) to 
subtract the required magnitude from that given for the parent in
the final catalogue, and use the colour transforms to allow for any differences
in instrumental magnitude introduced by both stars' colours.  

If we simply added the scaled, sky subtracted parent to the image at the
position of the required ghost the final image would not have the
correct noise properties.
For example, the noise in the wings of the ghost should correspond to the
noise in the sky at this point, and so the addition of a noiseless parent
would have the correct effect.
However, the parents are never noiseless, and so without some further 
procedure one would always have too much noise in the ghosts' wings.
Before adding the ghost, therefore, we replace each pixel in the area with
the median of itself and the eight pixels around it, removing the majority 
of the sky noise.
We then calculate the noise required in the ghost, subtract from it the noise
already present in the scaled parent, and then perturb the final image by
this amount.
This, of course, requires that the parents are always considerably brighter
than the ghosts.
The median process will also slightly broaden any star the ghost lies close 
to, but this should not affect our completeness calculation.

It is important that we correctly simulate stars which lie in overlap areas
between two or more fields.
Their multiple measurements are averaged, resulting in a higher
signal-to-noise ratio for a given magnitude, which can affect any
completeness which includes a signal-to-noise ratio cutoff.
Our creation of the ghost and parent list is therefore carried out from the 
final catalogue.
The celestial co-ordinates of every star are then translated back to pixel
co-ordinates for every image, and the star inserted if it lies on the image.
The only potential problem here would be if a ghost lay on an image, but not
its parent.
To ensure this does not happen, the positional shift from parent to ghost is
always parallel to the closest CCD edge and towards the middle of that edge.

Once the simulated stars have been introduced into each image of a given field,
it is straightforward to run these images through the entire data reduction
process. 
In practice it is much faster to carry out only photometry of the stars
which have been injected into the images.
Thus, after the star detection algorithm has run, we select those stars
which lie close to the position of ghosts, and only carry out photometry of
these objects.

\subsection{Testing the technique}

We can use the area of the wide catalogue which is covered by the deep 
catalogue to {\it measure} the completeness function for the wide catalogue, 
and then compare it with the {\it prediction} made by our completeness 
correction.
Although such a test is clearly crucial to validating any correction, we are
unaware of similar experiments being carried out to test other techniques.

\begin{figure}
\vspace{70mm}
\includegraphics{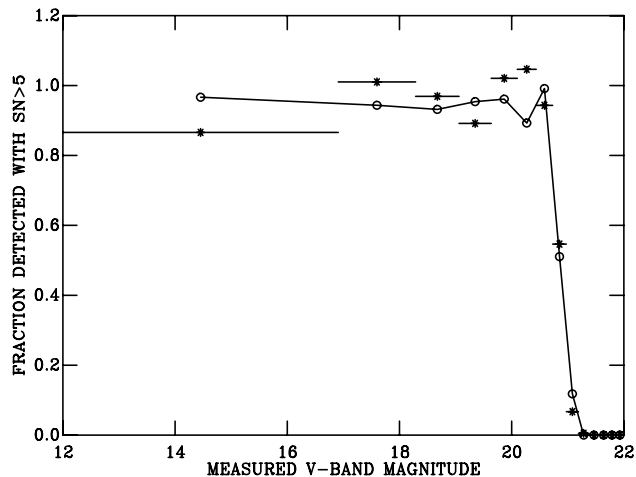}
\caption{
The completeness function (including magnitude creep) for the area
of the wide survey which overlaps with the deep survey.
The measured completeness function is the line with circles,
that predicted by simulation is marked as bars with stars.
The magnitude bins for the two datasets are identical, and shown by 
the bars on the starred points.
}
\label{comp}
\end{figure}

In principle, to measure the completeness of the wide catalogue, one simply 
takes each star in the deep catalogue, and asks whether it was detected in 
the wide survey.
In practice, we know colour effects are important, so we began by selecting 
just those objects in the deep catalogue with colours $0.95 < V-I_c < 1.25$.
We then searched for stars with matching positions in the wide catalogue,
which also had flags of zero, and a signal-to-noise ratio greater than five.
The fraction of deep catalogue objects which appear 
in the wide catalogue and pass the above tests, is our measured  
completeness function.
It is plotted as a circled line in Fig. \ref{comp}.

We matched this by simulating, in the wide survey, a sequence with 
$V-I_c=1.1$ and $B-V=0.7$ (the $B-V$ is close to the mean for the dataset).
We increased the number of stars in the sequence exponentially with magnitude,
so that the number per magnitude increases by a factor 1.71 per magnitude;
similar to the increase with magnitude for the entire catalogue.
We took the ratio of detected simulated objects with flags of zero and a
signal-to-noise ratio greater than five, to input objects.  
The result as, a function of detected magnitude, is our predicted completeness 
function for the wide survey, plotted as stars in Fig. \ref{comp}.

Fig. \ref{comp} shows a remarkable degree of agreement between the
predictions of our simulations, and the actual completeness function.
Equally encouraging, the detection efficiency falls from 90 to 10 percent in 
about half a magnitude.
Such a sharp cut-off is highly desirable for constructing luminosity 
functions, since it limits the region affected by (possibly) inaccurate 
completeness corrections.
Finally we note that sometimes the completeness correction is greater than
one.
This ``magnitude creep'' is caused by objects being detected in a magnitude
bin other than that which they were simulated, either because of statistical
uncertainty or the presence of another star \citep[see][]{stetson91}.
In practice the nett effect of magnitude creep is to place too many stars
in the brighter bins, but the effect depends on the slope of the luminosity
function, so it is important the simulated one chosen is close to that of the 
data.
There is a further reason for matching the slope of the luminosity function in
the simulations. 
The completeness function is presented in bins where, close to the cut-off, 
both the luminosity function and the completeness function vary significantly
within the binning.
As long as the real and simulated luminosity function are similar, binned
data will be directly comparable.

\section{THE CLUSTER LUMINOSITY FUNCTION}
\label{lf}

\begin{figure}
\vspace{110mm}
\includegraphics{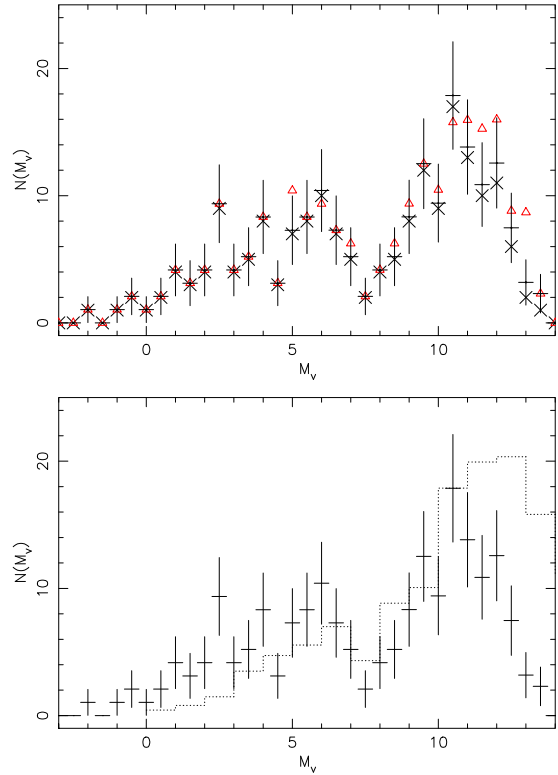}
\caption{
Top: The derived luminosity functions.
The functions derived using members selected with the D'Antona \&
Mazzitelli (1997) isochrones are show both with and without
completeness correction (error bars and diagonal crosses respectively).
The completeness-corrected function using the Siess (2000) selected members 
are marked by triangles.
Bottom: The corrected NGC 2547 luminosity function (derived using the 
D'Antona \& Mazzitelli isochrones) vs that for the field. 
The luminosity functions have been normalized for 8$<M_V<$10.
}
\label{lfplot}
\end{figure}

The objects selected as cluster candidates are placed into $V$
magnitude bins and plotted in Fig.~\ref{lfplot} as the cluster luminosity
function. 
The technique described in Section \ref{ghosts} was used to estimate the
correction for completeness.
A simulated PMS was used which matched the colours of the fitted PMS, with a 
similar growth in number with magnitude.
The derived correction by which the number of sources must be divided ($C$) 
was fitted with a function of the form
\begin{equation}
       C = B - A e^{{m-O}\over{W}},
\end{equation}
where $A$, $B$, $O$ and $W$ are constants, and $m$ is the magnitude referred
to.
Although in principle this correction can become negative, in practice our
there are no sources with an implied correction less than or equal to zero.
% A = 0.4796527
% M = 21.4891
% W = 0.7018
% B = 0.96085
This correction was applied on a star-by-star basis before binning, so that 
each star contributes a little more than 1 to the bin that it is placed in. 
Both corrected and uncorrected luminosity
functions are shown for comparison, as well as corrected luminosity
functions based on both the D'Antona \& Mazzitelli and the Siess et
al. isochrones. In Fig.~\ref{lfplot}, the x-axis has been transformed
to absolute $V$ magnitude by subtracting 8.25 (distance plus
extinction) from the observed magnitudes.

We believe that there is strong evidence for a turnover in the
luminosity function for $M_{v}>12$, although there are some reasons to
be cautious. The difference at faint magnitudes between the results
based on the D'Antona \& Mazzitelli and Siess et al. isochrones
illustrates that there is some model dependency. The Siess et
al. isochrone is fainter than the D'Antona \& Mazzitelli isochrone in
the $V$ vs $V-I_{\rm c}$ CMD at $V-I_{\rm c}\simeq3$, and as a result
an extra 15 faint stars are identified as candidate members.  However,
it is unrealistic to suppose that the steep decline seen even for the
Siess et al. luminosity function at $M_{v}>12$ could be reversed by a
model isochrone that is even fainter at red colours. For one thing, one
would start to lose candidate members from above the original
isochrones and for another, there would be a distinct possibility of
heavy contamination of the selected candidates by field stars, thus
increasing the actual decline in the luminosity function.

The second feature of the luminosity functions worth commenting on is
the presence of a strong ``Wielen dip'' at $7<M_{v}<8$ \citep{wielen74,
upgren81, bahcall86}.  This feature has been seen in several other open
clusters and the field \citep[see][]{jth01}, but is the first time it
has been seen in such a very young cluster.  In NGC 2547 the appearance
of the dip is enhanced by the likelihood of contaminating field objects
between $5.75<M_{v}<7.0$, but it is most certainly real as can be
judged from the scarcity of objects seen in the $V$ vs $V-I_{\rm c}$
CMD between $1.4<V-I_{\rm c}<1.7$. The dip is most likely caused by a
change in slope of the mass-luminosity relation, rather than a
discontinuity in the mass function (see below).

In Fig.\ref{lfplot}b we compare the completeness-corrected luminosity
function of NGC 2547 (derived from the D'Antona \& Mazzitelli
isochrones) with a normalized field star luminosity function
\citep[obtained from][]{reidhawley00}.  In a young cluster like NGC
2547 we would expect there to be relatively more short-lived, bright,
high mass stars, so we normalise the field luminosity function to NGC
2547 at $8<M_{V}<10$. The expected excess of bright stars at $M_{V}<4$
is indeed present in NGC 2547, but there also seems to be a deficit of
faint stars at $M_{v}>10$. The deficit is only marginally reduced if
the Siess et al isochrones are used instead, but may actually be
worsened if there are interloping non-members in the faint NGC 2547
sample. Of course, the steep decline in the NGC 2547 luminosity
function may not be the signature of a falling {\em mass} function. The
shape of the relationship between mass and luminosity (and $M_{v}$)
will be different from that for field stars in a very young cluster
like NGC 2547.

\section{THE MASS FUNCTION}
\label{func}

The cluster mass function (MF) is estimated by calculating the mass of
each candidate member using a mass-$V$ relationship derived from the
isochrones that were used to fit the cluster CMDs. We define the MF
using the form $dN/d\log{M} \propto M^{-\alpha}$, where the canonical
initial mass function (IMF) for field stars derived by
\cite{salpeter55} would have $\alpha=+1.35$.  Unresolved binarity is a
problem for this technique in two ways: A binary star that appears
above the single star cluster locus will be treated as one star with a
slightly higher mass than either of its two components. The second
problem is that binary stars with unequal mass components ($q<0.5-0.6$)
will have a derived primary star mass that is approximately correct,
but the hidden, lower mass secondary component will not be included in
the mass function. \cite{sagar91} and \cite{kroupa01} show that the
combination of these two effects could result in serious systematic
underestimates of $\alpha$ by between 0.3 and 1, depending on the mass
range considered, the binary fraction and the steepness of the mass
function.

We take a simple approach which addresses the first of these
difficulties. We {\em add} 0.5 mag to the $V$ of an object identified
as a $q\geq0.5$ binary ($q\geq0.6$ for $V<14$) in the candidate member
selection procedure and then put two stars with a mass appropriate for
this new magnitude into the corresponding mass function bin. The second
problem is less easy to deal with because we do not know what the
binary fraction or distribution of mass ratios are as a function of
mass. We estimate the possible effects by simulating a binary fraction
and mass ratio distribution among those stars considered single (see
below). Attempting to correct the observed mass function for binarity
is important if one wants to calculate a total mass or estimate the
true mass function. It is not an issue when comparing one cluster with
another or with the field, so long as corrections have been applied in
the same way (or not at all) and there are no significant differences
in binary fraction or mass ratio distribution. There is only marginal
evidence for an enhanced binary fraction in younger clusters compared
to the field \citep{bouvier01}, but we know little about the evolution
of mass ratio distributions. In field stars and the Pleiades, $q$ has a relatively flat
distribution between 0.1 and 1, although there is some evidence for a
small rise in frequency at $q\simeq0.3$ \citep{duquennoy91, kahler99}.
In this paper we will quote and display as our main results the mass functions after
applying our correction for binaries with large $q$. We choose to do
this because it is a simple scheme that could easily be applied to any
observational dataset. We will however examine the effects of not
including this correction or attempting to include a simulated binary
population with small $q$.

\begin{figure}
\vspace{70mm}
\includegraphics{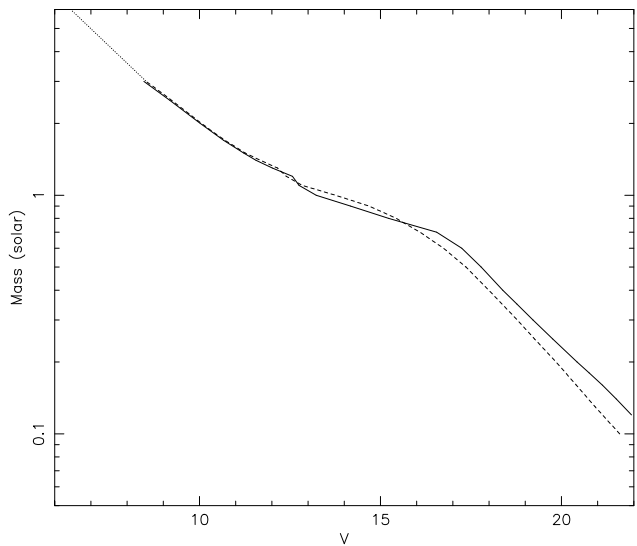}
\caption{
The adopted mass vs $V$-magnitude relationships.
The dashed line is from Siess (2000), the solid line from  D'Antona \&
Mazzitelli (1997), and 
the dotted line from Schaller (1992).
}
\label{massvplot}
\end{figure}

We show the mass-$V$ relations used in Fig. \ref{massvplot}. The mass-$V$
relationships from the isochrones are only well calibrated to
3$M_{\odot}$. This is the maximum mass in the D'Antona \& Mazzitelli
models, but it is also approximately the point at which $B-V$ becomes
bluer than the available Pleiades objects used to calibrate the
colour-$T_{\rm eff}$ relation. For the four stars with higher masses we
have used a mass-$V$ relationship from the models of \cite{schaller92}, 
which is shown as a dotted line in Fig.~\ref{massvplot}.

\begin{figure}
\vspace{70mm}
\includegraphics{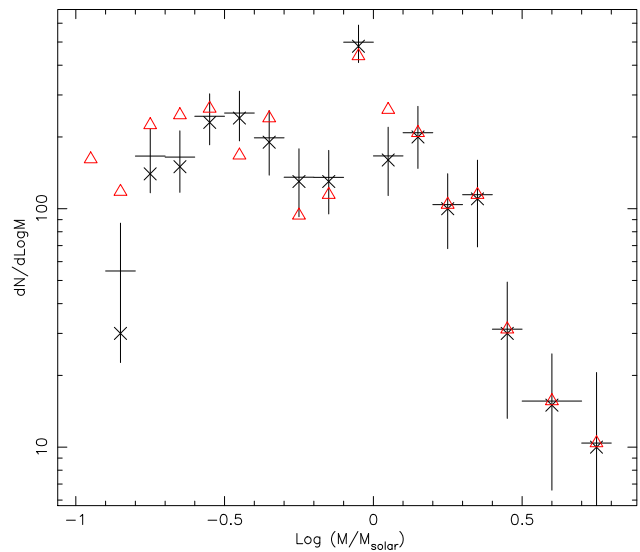}
\caption{
The derived mass functions.
The triangles use the models of Siess (2000) the crosses those of  D'Antona \&
Mazzitelli (1997); neither are corrected for completeness.
The error bars are the points derived from D'Antona \&
Mazzitelli (1997), corrected for completeness.
}
\label{mfdam97plot}
\end{figure}

The derived MFs are shown in Fig.~\ref{mfdam97plot}. MFs that are both
corrected and uncorrected for completeness are shown for the D'Antona
\& Mazzitelli isochrones, so that the reader can judge what effect
these corrections have. The corrections have an almost negligible
effect on the MF shape until the very lowest masses. The effects of
field star contamination in our sample will be seen as an overestimate
of the MF. For $14.0<V<15.25$, where we expect the contamination could
be as high as 40 percent, this corresponds to a mass range of
$0.8<M<1.0M_{\odot}$. This contamination is therefore confined to the
one high point in the MF at $\log(M/M_{\odot})=-0.05\pm0.05$, which we
ignore when fitting the mass function (see below).

At slightly fainter magnitudes and lower masses, we see that the
shallower gradient in the D'Antona \& Mazzitelli mass-$V$ relation at
$V\sim16$ has almost completely removed the ``Wielen dip'' that was
seen in the luminosity function. 
The same cannot be said for the Siess et al. mass-$V$
relation and derived MF, where there still appears to be a significant
depression for $-0.3<\log(M/M_{\odot})<-0.1$.

If we were to consider the evidence from the D'Antona \& Mazzitelli
results alone, we might have concluded that there were indications that
the MF turns downward at $M<0.2M_{\odot}$. Comparison with the results
from the Siess et al. models show that such a conclusion would be
premature. There are two basic reasons for the discrepant behaviour at
low masses. First, the Siess et al. isochrones follow a somewhat lower
path (by $\sim 0.1$ mag) in the $V$ vs $V-I_{\rm c}$ CMD and as a
result there are an extra 15 stars included in the luminosity function at faint
magnitudes (see Fig.\ref{lfplot}). Second, the mass-$V$ relationship
for the two sets of models are similar above 1$M_{\odot}$, but become
significantly different at low masses. The Siess et al. models yield a
lower mass for the same $V$ magnitude, which has the effect of moving
stars into lower mass bins when compared with the D'Antona \&
Mazzitelli MF.

There are additional reasons to be cautious in interpreting the
low-mass MF, not connected with the chosen evolutionary models.  Our
photometric calibration at $V-I_{\rm c}$ is not tightly constrained for
very red stars. Only three standards were measured with $V-I_{\rm c}>2$
(the reddest of which had $V-I_{\rm c}=2.7$), and although these stars
are well fitted with a colour term very close to unity (i.e. the filter
combination almost perfectly matches the Cousins system), we have to
realize that the colours and magnitudes for redder stars are based on
an extrapolated calibration. This colour limit corresponds to
approximately $V\simeq18.5$ in the cluster, or masses of about
0.35$M_{\odot}$. 
A change in the colour calibration beyond this point
could result in a different number of stars being selected as members
of the cluster (because the isochrone shape would change).
It would also have the more
minor effect of changing the deduced masses for the cluster candidates, if
their $V$ magnitudes were significantly altered. However, the
appearance of the CMD gives us some confidence that this is not likely
to yield MFs outside of the range encompassed by the two sets of
evolutionary models that we have used here. 

The MFs we have derived show the same general features that have been
seen in the field and in other young cluster like the Pleiades, with a
steep increase from high to low masses which flattens below 1$M_{\odot}$. 
We have
parameterized the MFs by fitting power laws of the form $dN/d\log{M}
\propto M^{-\alpha}$ to the sections from $0.1<M<0.8M_{\odot}$
(stopping at 0.16$M_{\odot}$ for the D'Antona \& Mazzitelli derived
MFs) and $1<M<6M_{\odot}$. We find $\alpha = 0.20\pm0.25$ or
$0.20\pm0.18$ for the low-mass stars using the D'Antona \& Mazzitelli
or Siess et al. models respectively, and $\alpha = 1.8\pm0.4$ or
$2.0\pm0.4$ for the high-mass stars.

The results for the higher mass stars seem a little steeper than the
$\alpha=1.40\pm0.13$ found for intermediate mass stars in many open
clusters by \cite{phelps93} and also than the canonical field MF
value of $\alpha=1.35$ found by \cite{salpeter55}. Some of the difference
can be attributed to our technique of splitting the $q>0.6$ unresolved
binary systems into two stars of slightly lower masses. This practice
was not adopted by the cited papers, which used a single mass-$V$
relation irrespective of binary status. This would not greatly affect
the results if binary fraction were independent of mass {\em and} there
were no high-mass cut-off to the distribution. However this is not the
case in our data, which is limited to $M<6M_{\odot}$. We therefore find
that if we were to adopt the same procedure as the previous work, our
MFs would be less steep and $\alpha$ reduced to $1.3\pm0.4$.

The results for low-mass stars are very similar to those derived for
the Pleiades \citep[where $\alpha\simeq 0$ for $0.3<M<1.0M_{\odot}$
--][]{Meusinger96}, M35 \citep[where $\alpha=-0.2$ for
$0.2<M<0.8M_{\odot}$ -- ][]{barrado01} and the field \citep[where
$-0.1<\alpha<+0.3$ for $0.1<M<0.6M_{\odot}$ --][] {gould97, kroupa01}.
We find that our treatment of $q>0.5$ binary systems has a small
effect. If we had not used this correction, our $\alpha$ would be smaller
by 0.2 and thus approximately zero.

The total mass of the cluster (inside the deep survey area) can now be
estimated by summing the masses of the individual stellar systems. A
simple integration of the mass function yields 190$M_{\odot}$ down to
stellar masses of $0.1M_{\odot}$ (irrespective of which evolutionary
model is chosen).  This figure contains a partial correction for those
binary systems with $q>0.5$, which accounts for $30\pm4$ percent of the
total number of stellar systems. This binary fraction is a little
higher than for equivalent systems in G-type field stars \citep[22
percent --][] {duquennoy91}, but very similar to the 26 percent found
by similar techniques among the G and K stars of the Pleiades and NGC
2516 \citep{stauffer84, jth01}. There is some evidence that the binary
fraction among lower mass field stars is smaller (30-40 percent), but
that they are more inclined to be found in high mass ratio systems
\citep{fisher92}.  If we were to assume a flat $q$ distribution down to
$q=0$, then
our measurements are consistent with a total binary fraction of about
60 percent for NGC 2547. Adopting the \cite{duquennoy91} field star $q$
distribution, we would estimate a total binary fraction of
about 90 percent.

To account for these hidden binary companions in $q<0.5$ systems, or at
least illustrate what effects they might have on the mass function and
total cluster mass, we simulate the unobserved binary population
assuming a total binary fraction and mass ratio distribution.  We
assume a total binary fraction of 90 percent and then randomly randomly
allocate lower mass binary companions to the ``single'' stars, with $q$
drawn from the distribution proposed by \citep{duquennoy91}. We
generate a mass function in the same way as previously and integrate
this function to get the total mass (above $0.1M_{\odot}$). The mass
function is only slightly changed from those discussed above. $\alpha$
is increased by about 0.1 for the low mass stars and unchanged for the
high mass stars.  The total mass of the cluster increases to about
240$M_{\odot}$.

To compare the mass of NGC 2547 with other clusters we need also to
make a correction for the fact that our deep survey area only covers a
fraction of the cluster. To estimate this fraction we simply note that
Clari\'{a}'s (1982) photometry of bright stars in the region appears to
be complete to $V=12$ and covered a much larger area (50 arcmin radius)
than our own. We find that 65 percent of Clari\'{a}'s proposed cluster
members with $V<12$ (which almost entirely agree with our own
classifications where there is overlap) lie within the bounds of our
deep survey area. Assuming that this fraction is mass-independent,
i.e. that there is no mass segregation in this young cluster, we obtain
a final estimate of the cluster mass of 370$M_{\odot}$. This is
considerably less (by factors of 3-5) than rich open clusters like the
Pleiades, M35 and NGC 2516 (e.g. Meusinger et al. 1996; Barrado y
Navascu\'{e}s et al. 2001; Jeffries et al. 2001). Thus an important
result is that the mass function of NGC 2547 is very similar to that of
clusters with much larger masses, providing some support for the idea
that the initial mass function is universal and independent of
environment \citep{kroupa02}.

\section{CONCLUSIONS}
\label{conc}

In this paper we have developed a set of algorithms for effectively
dealing with the analysis of large CCD, multi-colour mosaics, with the
ultimate aim of producing colour-magnitude diagrams with
well-understood completeness properties.
We have shown that the uncertainty estimate from these techniques is robust,
and that the resulting colour-magnitude diagrams reach approximately 0.45 
magnitudes deeper than the same dataset reduced using aperture photometry.
We have also shown that the completeness correction technique is reliable.

To obtain the best from the software (or indeed any similar data reduction 
procedure), we recommend the following observing strategy.
\\
\\
1) The area to be surveyed should be divided into fields, with an overlap
between each field.
The reasons for this are given in Section \ref{fcal}, but in summary are
as follows.  
\hfil\break
(i)Minimizing the difference between the measurements of the same stars in
different fields improves the self-consistency of the catalogue.
\hfil\break
(ii) The remaining differences give an estimate of the systematic calibration 
errors.
\\
\\
2) If magnitudes or positions of bright stars are required (for example
as fiducial stars for follow-up fibre spectroscopy), a single short exposure
should be made for each field in each band.
\\
\\
3) The signal-to-noise ratio for the faintest stars should then be built up through 
at least three longer exposures.
There are two reasons for preferring this to a single, longer exposure.
\hfil\break
(i) $\chi^2$ can be used to distinguish spurious events in a single frame from
objects which appear in all frames (Section \ref{comb_meas}).
\hfil\break
(ii) The final effective profile correction is a mean of several measurements,
yielding greater precision.
\\
\\
We have applied our techniques to the young open cluster, NGC
2547. From X-ray selected members of the cluster in the $B-V$,$V$ and
$V-I_{\rm c}$,$V$ colour-magnitude diagrams we have estimated a new
(model-dependent) intrinsic distance modulus of 8.00-8.15 and an age of
20-35\,Myr. The fitted isochrones have been used to photometrically
select cluster members and derive completeness-corrected luminosity and
mass functions between 0.1 and 6$M_{\odot}$.  The luminosity function
shows some evidence for a lack of stars with $M_{v}>10$ when compared
with the field and Pleiades functions. There is also strong evidence
for a ``Wielen dip'' at $M_{v}\sim 8$. We find that this luminosity
function structure is largely absent in the mass function, which shows
great similarity to the mass functions found in the field and other
open clusters. The total cluster mass (down to stars of 0.1$M_{\odot}$)
is about 190$M_{\odot}$, or 370$M_{\odot}$ if we make approximate
corrections for unresolved binaries and stars lying outside our survey
area.  Thus NGC 2547 is a factor of 3-5 less massive than some older
open clusters like the Pleiades, M35 and NGC 2516, yet has an extremely
similar mass function.

\section*{Acknowledgments}

We are grateful to PPARC for supporting TN through the advanced fellowship 
programme, SAT through the studentship programme and for research grant 
support for EJT. 
CRD was supported by a Nuffield Undergraduate Bursary (NUF-URB00).
We would like to thank the director and staff of the Cerro Tololo
Interamerican Observatory, operated
by the Association of Universities for Research in Astronomy, Inc.,
under contract to the US National Science Foundation.
Computing was performed
at the Keele and Exeter nodes of the Starlink network, funded by PPARC.

\bibliographystyle{mn2e}
\bibliography{text}

\begin{thebibliography}{}

\bibitem[\protect\citeauthoryear{{Alcock}, {et al} \& {}}{{Alcock}
  et~al.}{1996}]{MACHO}
{Alcock} C.,  {et al}   {} 1996, ApJ, 461, 84

\bibitem[\protect\citeauthoryear{{Arnouts}, {Vandame}, {Benoist},
  {Groenewegen}, {da Costa}, {Schirmer}, {Mignani}, {Slijkhuis},
  {Hatziminaoglou}, {Hook}, {Madejsky}, {Rit{\' e}} \& {Wicenec}}{{Arnouts}
  et~al.}{2001}]{eis}
{Arnouts} S.,  {Vandame} B.,  {Benoist} C.,  {Groenewegen} M.~A.~T.,  {da
  Costa} L.,  {Schirmer} M.,  {Mignani} R.~P.,  {Slijkhuis} R.,
  {Hatziminaoglou} E.,  {Hook} R.,  {Madejsky} R.,  {Rit{\' e}} C.,
  {Wicenec} A.,  2001, A\&A, 379, 740

\bibitem[\protect\citeauthoryear{Bahcall}{Bahcall}{1986}]{bahcall86}
Bahcall J.~N.,  1986, ARA\&A, 24, 577

\bibitem[\protect\citeauthoryear{{Barrado~y Navascu\'{e}s}, Stauffer, Bouvier
  \& {Mart\'{i}n}}{{Barrado~y Navascu\'{e}s} et~al.}{2001}]{barrado01}
{Barrado~y Navascu\'{e}s} D.,  Stauffer J.~R.,  Bouvier J.,    {Mart\'{i}n}
  E.~L.,  2001, ApJ, 546, 1006

\bibitem[\protect\citeauthoryear{Bertin \& Arnouts}{Bertin \&
  Arnouts}{1996}]{sextractor}
Bertin E.,  Arnouts S.,  1996, A\&AS, 117, 393

\bibitem[\protect\citeauthoryear{Bessell, Castelli \& Plez}{Bessell
  et~al.}{1998}]{bessell98}
Bessell M.~S.,  Castelli F.,    Plez B.,  1998, A\&A, 333, 231

\bibitem[\protect\citeauthoryear{Bouvier, Duch\^{e}ne, Mermilliod \&
  Simon}{Bouvier et~al.}{2001}]{bouvier01}
Bouvier J.,  Duch\^{e}ne G.,  Mermilliod J.~C.,    Simon T.,  2001, A\&A, 375,
  989

\bibitem[\protect\citeauthoryear{Bouvier, Rigaut \& Nadeau}{Bouvier
  et~al.}{1997}]{bouvier97bin}
Bouvier J.,  Rigaut F.,    Nadeau D.,  1997, A\&A, 323, 139

\bibitem[\protect\citeauthoryear{Caldwell, Keane \& Schechter}{Caldwell
  et~al.}{1991}]{caldwell91}
Caldwell J.,  Keane M.,    Schechter P.,  1991, AJ, 101

\bibitem[\protect\citeauthoryear{Clari\'a}{Clari\'a}{1982}]{claria82}
Clari\'a J.,  1982, A\&ASS, 47, 323

\bibitem[\protect\citeauthoryear{{D'Antona} \& Mazzitelli}{{D'Antona} \&
  Mazzitelli}{1997}]{dantona97}
{D'Antona} F.,  Mazzitelli I.,  1997, Mem. Soc. Astr. It., 68, 807

\bibitem[\protect\citeauthoryear{Duquennoy \& Mayor}{Duquennoy \&
  Mayor}{1991}]{duquennoy91}
Duquennoy A.,  Mayor M.,  1991, A\&A, 248, 485

\bibitem[\protect\citeauthoryear{Eaton, Draper \& Allan}{Eaton
  et~al.}{1999}]{photom}
Eaton N.,  Draper P.,    Allan A.,  1999, PHOTOM -- A Photometry Package
  version 1.8-3 Starlink User Note 45.9, Starlink Project

\bibitem[\protect\citeauthoryear{{Epchtein}, , {et al} \& {}}{{Epchtein}
  et~al.}{1997}]{denis}
{Epchtein} N.,   {et al}   {} 1997, The Messenger, 87, 27

\bibitem[\protect\citeauthoryear{{Fischer} \& {Kochanski}}{{Fischer} \&
  {Kochanski}}{1994}]{fischer94}
{Fischer} P.,  {Kochanski} G.~P.,  1994, AJ, 107, 802

\bibitem[\protect\citeauthoryear{Fisher \& Marcy}{Fisher \&
  Marcy}{1992}]{fisher92}
Fisher D.~A.,  Marcy G.~W.,  1992, ApJ, 396, 178

\bibitem[\protect\citeauthoryear{Flower}{Flower}{1996}]{flower96}
Flower P.~J.,  1996, ApJ, 469, 355

\bibitem[\protect\citeauthoryear{Gould, Bahcall \& Flynn}{Gould
  et~al.}{1997}]{gould97}
Gould A.,  Bahcall J.~N.,    Flynn C.,  1997, ApJ, 482, 913

\bibitem[\protect\citeauthoryear{{Hambly}, {Davenhall}, {Irwin} \&
  {MacGillivray}}{{Hambly} et~al.}{2001}]{sss3}
{Hambly} N.~C.,  {Davenhall} A.~C.,  {Irwin} M.~J.,    {MacGillivray} H.~T.,
  2001, MNRAS, 326, 1315

\bibitem[\protect\citeauthoryear{{Hambly}, {MacGillivray}, {Read}, {Tritton},
  {Thomson}, {Kelly}, {Morgan}, {Smith}, {Driver}, {Williamson}, {Parker},
  {Hawkins}, {Williams} \& {Lawrence}}{{Hambly} et~al.}{2001}]{sss1}
{Hambly} N.~C.,  {MacGillivray} H.~T.,  {Read} M.~A.,  {Tritton} S.~B.,
  {Thomson} E.~B.,  {Kelly} B.~D.,  {Morgan} D.~H.,  {Smith} R.~E.,  {Driver}
  S.~P.,  {Williamson} J.,  {Parker} Q.~A.,  {Hawkins} M.~R.~S.,  {Williams}
  P.~M.,    {Lawrence} A.,  2001, MNRAS, 326, 1279

\bibitem[\protect\citeauthoryear{Irwin}{Irwin}{1997}]{irwin97}
Irwin M.,  1997, in {Rodr\'iguez} J.,  Espinosa A.,  Herrero A.,   {S\'anchez}
  F.,  eds, Instrumentation for Large Telescopes (VII Canary Islands Winter
  School) Cambridge University Press

\bibitem[\protect\citeauthoryear{Jeffries, Thurston \& Hambly}{Jeffries
  et~al.}{2001}]{jth01}
Jeffries R.,  Thurston M.,    Hambly N.,  2001, A\&A, 375, 863

\bibitem[\protect\citeauthoryear{Jeffries \& Tolley}{Jeffries \&
  Tolley}{1998}]{jt98}
Jeffries R.,  Tolley A.,  1998, MNRAS, 300, 331

\bibitem[\protect\citeauthoryear{Jeffries, Totten \& James}{Jeffries
  et~al.}{2000}]{jtj00}
Jeffries R.,  Totten E.,    James D.,  2000, MNRAS, 316, 950

\bibitem[\protect\citeauthoryear{{K\"ahler}}{{K\"ahler}}{1999}]{kahler99}
{K\"ahler} H.,  1999, A\&A, 346, 67

\bibitem[\protect\citeauthoryear{Kalirai, Richer, Fahlman, Cuillandre, Ventura,
  {D'Antona}, Bertin, Marconi \& Durrell}{Kalirai et~al.}{2001}]{kalirai01}
Kalirai J.,  Richer H.,  Fahlman G.,  Cuillandre J.-C.,  Ventura P.,
  {D'Antona} F.,  Bertin E.,  Marconi G.,    Durrell P.,  2001, AJ, 122, 266

\bibitem[\protect\citeauthoryear{King}{King}{1983}]{king83}
King I.,  1983, PASP, 95, 163

\bibitem[\protect\citeauthoryear{Kroupa}{Kroupa}{2001}]{kroupa01}
Kroupa P.,  2001, MNRAS, 322, 231

\bibitem[\protect\citeauthoryear{Kroupa}{Kroupa}{2002}]{kroupa02}
Kroupa P.,  2002, Science, 295, 82

\bibitem[\protect\citeauthoryear{Leggett, Allard, Berriman, Dahn \&
  Hauschildt}{Leggett et~al.}{1996}]{leggett96}
Leggett S.~K.,  Allard F.,  Berriman G.,  Dahn C.~C.,    Hauschildt P.~H.,
  1996, ApJS, 104, 117

\bibitem[\protect\citeauthoryear{Mermilliod, Rosvick, Duquennoy \&
  Mayor}{Mermilliod et~al.}{1992}]{mermilliod92}
Mermilliod J.~C.,  Rosvick J.~M.,  Duquennoy A.,    Mayor M.,  1992, A\&A, 265,
  513

\bibitem[\protect\citeauthoryear{Meusinger, Schilbach \& Souchay}{Meusinger
  et~al.}{1996}]{Meusinger96}
Meusinger H.,  Schilbach E.,    Souchay J.,  1996, A\&A, 312, 833

\bibitem[\protect\citeauthoryear{{Momany}, {Vandame}, {Zaggia}, {Mignani}, {da
  Costa}, {Arnouts}, {Groenewegen}, {Hatziminaoglou}, {Madejsky}, {Rit{\' e}},
  {Schirmer} \& {Slijkhuis}}{{Momany} et~al.}{2001}]{pre-flames}
{Momany} Y.,  {Vandame} B.,  {Zaggia} S.,  {Mignani} R.~P.,  {da Costa} L.,
  {Arnouts} S.,  {Groenewegen} M.~A.~T.,  {Hatziminaoglou} E.,  {Madejsky} R.,
  {Rit{\' e}} C.,  {Schirmer} M.,    {Slijkhuis} R.,  2001, A\&A, 379, 436

\bibitem[\protect\citeauthoryear{Monet}{Monet}{1998}]{monet98}
Monet D.,  1998, BAAS, 30, 1427

\bibitem[\protect\citeauthoryear{{Mutchier} \& {Fruchter}}{{Mutchier} \&
  {Fruchter}}{1997}]{drizzle}
{Mutchier} M.,  {Fruchter} A.,  1997, in The 1997 HST Calibration Workshop with
  a new generation of instruments /edited by Stefano Casertano, Robert
  Jedrzejewski, Charles D. Keyes, and Mark Stevens. Baltimore, MD : Space
  Telescope Science Institute (1997) QB 500.268 C35 1997, p. 355. {Drizzling
  Dithered WFPC2 Images-A Demonstration}.
p.~355

\bibitem[\protect\citeauthoryear{Naylor}{Naylor}{1998}]{p1}
Naylor T.,  1998, MNRAS, 296, 339

\bibitem[\protect\citeauthoryear{Naylor, Jeffries \& Pozzo}{Naylor
  et~al.}{2001}]{naylor01}
Naylor T.,  Jeffries R.~D.,    Pozzo M.,  2001, in Montmerle T.,  Andre P.,
  eds, ASP Conf. Ser. 243: From Darkness to Light p.~705

\bibitem[\protect\citeauthoryear{Phelps \& Janes}{Phelps \&
  Janes}{1993}]{phelps93}
Phelps R.~L.,  Janes K.~A.,  1993, AJ, 106, 1870

\bibitem[\protect\citeauthoryear{{Pozzo}, {Jeffries}, {Naylor}, {Totten},
  {Harmer} \& {Kenyon}}{{Pozzo} et~al.}{2000}]{monica}
{Pozzo} M.,  {Jeffries} R.~D.,  {Naylor} T.,  {Totten} E.~J.,  {Harmer} S.,
  {Kenyon} M.,  2000, MNRAS, 313, L23

\bibitem[\protect\citeauthoryear{Reid \& Hawley}{Reid \&
  Hawley}{2000}]{reidhawley00}
Reid I.~N.,  Hawley S.~L.,  2000, in New light on dark stars Springer-Praxis

\bibitem[\protect\citeauthoryear{{Robichon}, {Arenou}, {Mermilliod} \&
  {Turon}}{{Robichon} et~al.}{1999}]{robichon99}
{Robichon} N.,  {Arenou} F.,  {Mermilliod} J.-C.,    {Turon} C.,  1999, A\&A,
  345, 471

\bibitem[\protect\citeauthoryear{Sagar \& Richtler}{Sagar \&
  Richtler}{1991}]{sagar91}
Sagar R.,  Richtler T.,  1991, A\&A, 250, 324

\bibitem[\protect\citeauthoryear{Salpeter}{Salpeter}{1955}]{salpeter55}
Salpeter E.~E.,  1955, ApJ, 121, 161

\bibitem[\protect\citeauthoryear{Schaller, Schaerer, Meynet \& Maeder}{Schaller
  et~al.}{1992}]{schaller92}
Schaller G.,  Schaerer D.,  Meynet G.,    Maeder A.,  1992, A\&AS, 96, 269

\bibitem[\protect\citeauthoryear{{Siess}, {Dufour} \& {Forestini}}{{Siess}
  et~al.}{2000}]{siess00}
{Siess} L.,  {Dufour} E.,    {Forestini} M.,  2000, A\&A, 358, 593

\bibitem[\protect\citeauthoryear{Stauffer, Hartmann, Soderblom \&
  Burnham}{Stauffer et~al.}{1984}]{stauffer84}
Stauffer J.~R.,  Hartmann L.,  Soderblom D.~R.,    Burnham N.,  1984, ApJ, 280,
  202

\bibitem[\protect\citeauthoryear{Stauffer, Schultz \& Kirkpatrick}{Stauffer
  et~al.}{1998}]{stauffer98}
Stauffer J.~R.,  Schultz G.,    Kirkpatrick J.~D.,  1998, ApJ, 499, L199

\bibitem[\protect\citeauthoryear{Stetson}{Stetson}{1987}]{daophot}
Stetson P.,  1987, PASP, 99

\bibitem[\protect\citeauthoryear{Stetson}{Stetson}{1991}]{stetson91}
Stetson P.,  1991, in Janes K.,  ed., ASP Conf. Ser. 13: The Formation and
  Evolution of Star Clusters pp 88--111

\bibitem[\protect\citeauthoryear{Stover \& Allen}{Stover \& Allen}{1987}]{sa}
Stover R.,  Allen S.,  1987, PASP, 99, 877

\bibitem[\protect\citeauthoryear{Totten, Jeffries \& Hambly}{Totten
  et~al.}{2000}]{totten00}
Totten E.~J.,  Jeffries R.~D.,    Hambly N.~C.,  2000, in Pallavicini R.,
  Micela G.,   Sciortino S.,  eds, ASP Conf. Ser. 198: Stellar clusters and
  associations: Rotation, activity and dynamos p.~55

\bibitem[\protect\citeauthoryear{{Udalski}, {Kubiak} \& {Szymanski}}{{Udalski}
  et~al.}{1997}]{ogle}
{Udalski} A.,  {Kubiak} M.,    {Szymanski} M.,  1997, Acta Astronomica, 47, 319

\bibitem[\protect\citeauthoryear{Upgren \& Armandroff}{Upgren \&
  Armandroff}{1981}]{upgren81}
Upgren A.~R.,  Armandroff T.~E.,  1981, AJ, 86, 1898

\bibitem[\protect\citeauthoryear{von Hippel \& Sarajedini}{von Hippel \&
  Sarajedini}{1998}]{hippel98}
von Hippel T.,  Sarajedini A.,  1998, ApJ, 116

\bibitem[\protect\citeauthoryear{Wallace}{Wallace}{1998}]{wallace98}
Wallace P.,  1998, ASTROM -- Basic astrometry program v3.6 -- User's Guide,
  Starlink User Note 5.17, Starlink Project

\bibitem[\protect\citeauthoryear{Wallace}{Wallace}{1999}]{wallace99}
Wallace P.,  1999, SLALIB -- Positional Astronomy Library 2.4-0 Programmer's
  Manual, Starlink User Note 67.45, Starlink Project

\bibitem[\protect\citeauthoryear{Wielen}{Wielen}{1974}]{wielen74}
Wielen R.,  1974, in Contopoulos G.,  ed., Highlights of Astronomy, Vol. 3
  Dordrecht: Reidel, p.~395

\bibitem[\protect\citeauthoryear{{York}, {et al} \& {}}{{York}
  et~al.}{2000}]{sloan}
{York} D.~G.,  {et al}   {} 2000, AJ, 120, 1579

\bibitem[\protect\citeauthoryear{Zacharias, Urban, Zacharias, Hall, G.L., T.J.,
  Germain, Holdenried, Pohlman, F.S., D.G. \& Winter}{Zacharias
  et~al.}{2000}]{zacharias00}
Zacharias N.,  Urban S.,  Zacharias M.,  Hall D.,  G.L. W.,  T.J. R.,  Germain
  M.,  Holdenried E.,  Pohlman J.,  F.S. G.,  D.G. M.,    Winter L.,  2000, AJ,
  120, 2131

\end{thebibliography}

\begin{landscape}
\textwidth=240mm
\begin{table*}
  \begin{tabular}{@{}ccccc..cccccccccc@{}}
            &       &      & 
\multicolumn{2}{c}{J2000 Position} & 
\multicolumn{2}{c}{Pixel Position} & 
\multicolumn{3}{c}{V} &
\multicolumn{3}{c}{B-V} &
\multicolumn{3}{c}{V-I} \\
Clari\'a ID & Field & Star &                  
R.A. & Dec. &                
X & Y &
Mag. & Error & Flag  & 
Mag. & Error & Flag  & 
Mag. & Error & Flag  \\ 
   2 &   13  &    3 &  08 10 20.53 & -49 14 13.05 &  936.482 & 1186.031 &    10.878   &   0.022  & 44 &  -1.139  &    0.027 &  44  & 2.270 &     0.030 &  44  \\
   4 &    4  &    9 &  08 10 16.08 & -49 02  5.58 &   44.039 &  220.261 &    10.671   &   0.005  & 44 &   0.401  &    0.007 &  44  &  0.837 &     0.008 &  44 \\
   5  &   1   &  14 &  08 10 59.55 & -49 17  2.86 &  550.354 &  861.607 &    11.107   &   0.005  & 44 &   0.473  &    0.007 &  44  &  0.975  &    0.008 &  44 \\
   6 &   13  &   12 &  08 10 27.20 & -49 09 51.51 &  774.996 &  533.727 &    10.926   &   0.022  & 44 &  -1.102  &    0.027 &  44  &  2.099  &    0.030 &  44 \\
   7 &    2  &   10 &  08 09 59.20 & -49 16 12.62 &  464.725 &  729.392 &    10.864   &   0.005  & 44 &   0.403  &    0.007 &  44  &  1.041  &    0.008 &  44 \\
   8 &    7  &   10 &  08 08 41.61 & -49 29 49.76 & 1775.237 & 1948.361 &     9.416   &   0.022  & 44 &   0.184  &    0.027 &  44  &  0.752  &    0.030 &  44 \\
   9 &    2  &   11 &  08 10  7.84 & -49 16 36.84 &  254.033 &  789.545 &    10.817   &   0.005  & 44 &   0.403  &    0.007 &  44  &  1.039  &    0.008 &  44 \\
  12 &   14  &   10 &  08 09 23.65 & -49 10 46.30 &  752.246 &  666.993 &     9.688   &   0.022  & 44 &  -0.039  &    0.027 &  44  &  0.681  &    0.030 &  44 \\
  13 &   13  &   15 &  08 09 56.01 & -49 19 30.08 & 1532.496 & 1978.130 &    10.889   &   0.022  & 44 &  -1.254  &    0.027 &  44  &  2.359  &    0.030 &  44 \\
  14 &   14  &    3 &  08 09 36.66 & -49 11 35.75 &  434.076 &  789.798 &    13.221   &   0.022  & 44 &  -0.635  &    0.027 &  44  & -0.477  &    0.031 &  44 \\
  15 &    4  &   13 &  08 09 57.52 & -49 08 20.64 &  497.708 & 1155.391 &    10.922   &   0.005  & 44 &   0.387  &    0.007 &  44  &  0.844  &    0.008 &  44 \\
  16 &   14  &   11 &  08 09 25.99 & -49 11 55.83 &  694.830 &  840.222 &     9.812   &   0.022  & 44 &  -0.148  &    0.027 &  44  &  0.463  &    0.030 &  44 \\
  18 &   18  &   13 &  08 10 20.82 & -49 03 37.47 &  940.175 & 1165.762 &     9.836   &   0.022  & 44 &   0.056  &    0.027 &  44  &  0.264  &    0.030 &  44 \\
  19 &   14  &   12 &  08 09 51.27 & -49 11 15.83 &   77.055 &  739.852 &   120.528   & 101.958  & 88 &  -0.321  &  130.247 &  88  &  0.776  &  144.121 &  88 \\
  21 &   14  &   13 &  08 08 49.85 & -49 13 43.98 & 1577.135 & 1112.282 &     9.770   &   0.022  & 44 &  -0.043  &    0.027 &  44  &  0.416  &    0.030 &  44 \\
  22 &   13  &   20 &  08 10 31.42 & -49 06 31.07 &  672.533 &   33.895 &    10.951   &   0.022  & 44 &  -0.935  &    0.027 &  44  &  1.487  &    0.030 &  40 \\
  23 &   17  &   13 &  08 09 49.49 & -49 01 49.55 &  131.648 &  895.277 &   120.693   & 101.958  & 88 &  -0.312  &  130.247 &  88  &  0.945  &  144.121 &  88 \\
  24 &    7  &   12 &  08 09 28.12 & -49 25 57.69 &  646.604 & 1366.610 &    10.011   &   0.022  & 04 &   0.650  &    0.027 &  04  &  1.060  &    0.030 &  44 \\
  25 &    7  &   11 &  08 09 49.11 & -49 25  4.62 &  136.097 & 1233.820 &   120.563   & 101.958  & 88 &  -0.058  &  130.247 &  88  &  0.782  &  144.121 &  88 \\
  26 &   14  &   14 &  08 09 52.16 & -49 11  2.61 &   55.325 &  706.881 &   120.510   & 101.958  & 88 &  -0.322  &  130.247 &  88  &  0.771  &  144.121 &  88 \\
  27 &   14  &   15 &  08 08 53.12 & -49 13 49.58 & 1497.047 & 1125.950 &     9.940   &   0.022  & 44 &  -0.064  &    0.027 &  44  &  0.282  &    0.030 &  40 \\
  28 &   18  &   16 &  08 10 42.32 & -48 57 25.73 &  414.249 &  238.239 &     9.859   &   0.022  & 44 &   0.070  &    0.027 &  44  &  0.113  &    0.030 &  40 \\
  30 &    4  &   18 &  08 10  8.41 & -49 00 43.71 &  232.312 &   16.317 &    10.747   &   0.005  & 44 &   0.365  &    0.007 &  44  &  0.710  &    0.008 &  44 \\
  31 &    2  &   14 &  08 10  6.04 & -49 14 18.47 &  298.178 &  444.686 &    10.938   &   0.005  & 44 &   0.369  &    0.007 &  44  &  0.774  &    0.008 &  44 \\
  32 &   12  &   13 &  08 11 25.84 & -49 12 29.22 &  909.568 &  924.451 &     9.994   &   0.022  & 44 &   0.087  &    0.027 &  44  &  0.185  &    0.030 &  40 \\
  33 &   19  &   11 &  08 11  8.62 & -49 00 16.90 & 1345.094 &  666.306 &    10.209   &   0.022  & 44 &   0.142  &    0.027 &  44  & -0.018  &    0.030 &  44 \\
  35 &    7  &   13 &  08 09 30.49 & -49 21 56.22 &  589.893 &  764.561 &    10.257   &   0.022  & 44 &   0.026  &    0.027 &  44  &  0.282  &    0.030 &  40 \\
  36 &    8  &   13 &  08 10 35.94 & -49 18 21.46 &  569.289 &  228.736 &    10.021   &   0.022  & 40 &   0.257  &    0.027 &  40  &  0.329  &    0.030 &  00 \\
  38 &   12  &   14 &  08 10 48.73 & -49 06 51.81 & 1819.852 &   86.366 &    10.103   &   0.016  & 00 &   0.376  &    0.019 &  00  &  0.489  &    0.021 &  00 \\
  39 &   17  &   14 &  08 09 29.56 & -49 06 16.79 &  619.150 & 1561.870 &     8.084   &   0.022  & 88 &   1.942  &    0.027 &  88  &  0.191  &    0.030 &  88 \\
  41 &    2  &   17 &  08 09 50.64 & -49 12 49.88 &  674.479 &  224.332 &    11.002   &   0.005  & 44 &   0.376  &    0.007 &  44  &  0.730  &    0.008 &  44 \\
  42 &   18  &   19 &  08 10 38.97 & -49 01 48.83 &  495.654 &  894.172 &    10.338   &   0.022  & 44 &  -0.100  &    0.027 &  44  & -0.042  &    0.030 &  40 \\
  43 &    2  &   16 &  08 10 13.50 & -49 19 13.23 &  115.607 & 1179.280 &    11.015   &   0.005  & 44 &   0.405  &    0.007 &  44  &  0.927  &    0.008 &  44 \\
  45 &    7  &   14 &  08 09 38.12 & -49 18 40.96 &  404.576 &  277.555 &    10.802   &   0.013  & 00 &   0.259  &    0.016 &  00  &  0.307  &    0.017 &  00 \\
  48 &   18  &   21 &  08 10 26.68 & -49 06 53.48 &  795.559 & 1654.082 &    10.521   &   0.022  & 00 &   0.213  &    0.027 &  00  &  0.287  &    0.030 &  00 \\
  49 &    2  &   19 &  08 09 50.58 & -49 13 19.90 &  675.937 &  299.162 &    11.128   &   0.005  & 44 &   0.362  &    0.007 &  44  &  0.668  &    0.008 &  44 \\
  50 &    7  &   15 &  08 08 45.74 & -49 23 47.67 & 1678.509 & 1045.302 &    10.728   &   0.022  & 00 &   0.292  &    0.027 &  00  &  0.328  &    0.030 &  00 \\
  51 &   14  &   17 &  08 09 46.11 & -49 14 27.08 &  202.980 & 1216.712 &    10.680   &   0.022  & 00 &   0.248  &    0.027 &  00  &  0.344  &    0.030 &  00 \\
  52 &   18  &   22 &  08 09 51.93 & -49 00 25.36 & 1649.861 &  689.060 &    10.715   &   0.022  & 00 &   0.143  &    0.027 &  00  &  0.160  &    0.030 &  00 \\
  53 &   13  &   28 &  08 10 31.52 & -49 16 43.80 &  667.636 & 1561.296 &    11.332   &   0.022  & 04 &   0.526  &    0.027 &  04  &  5.883  &    0.030 &  48 \\
  54 &    9  &   12 &  08 10 46.62 & -49 17 31.53 & 1875.384 &  108.580 &    10.760   &   0.013  & 00 &   0.442  &    0.016 &  00  &  0.540  &    0.017 &  00 \\
  56 &    7  &   16 &  08 09 35.57 & -49 27  1.55 &  465.125 & 1525.564 &    10.777   &   0.022  & 00 &   0.440  &    0.027 &  00  &  0.558  &    0.021 &  00 \\
  57 &    9  &   13 &  08 11 37.83 & -49 29 36.49 &  623.488 & 1911.992 &     6.388   &   0.022  & 88 &   2.308  &    0.027 &  88  &  1.045  &    0.030 &  88 \\
  58 &   19  &   13 &  08 11  3.25 & -49 00 37.78 & 1476.825 &  718.776 &    10.881   &   0.022  & 00 &   0.283  &    0.027 &  00  &  0.356  &    0.030 &  00 \\
  60 &   13  & 3728 &  08 09 48.31 & -49 09 38.19 & 1725.986 &  503.403 &    18.503   &   0.030  & 11 &   1.025  &    0.143 &  11  &  0.610  &    0.079 &  11 \\
  61 &    8  &   17 &  08 10 24.41 & -49 23 17.45 &  848.941 &  967.079 &    10.889   &   0.022  & 00 &   0.198  &    0.027 &  00  &  0.300  &    0.030 &  00 \\
  62 &   14  &   20 &  08 09 26.69 & -49 14 37.30 &  676.996 & 1242.751 &    10.887   &   0.022  & 00 &   0.306  &    0.027 &  00  &  0.367  &    0.030 &  00 \\
  63 &   13  &   33 &  08 10 27.71 & -49 12 10.01 &  761.754 &  878.948 &    11.353   &   0.022  & 04 &   0.226  &    0.024 &  04  &  0.863  &    0.026 &  40 \\
  64 &    2  &   23 &  08 10 16.70 & -49 15 17.62 &   37.895 &  591.971 &    11.155   &   0.005  & 44 &   0.320  &    0.007 &  44  &  0.554  &    0.008 &  44 \\
  65 &    7  &   17 &  08 09 30.51 & -49 20 44.74 &  589.619 &  586.347 &    11.081   &   0.022  & 00 &   0.356  &    0.027 &  00  &  0.427  &    0.030 &  00 \\
  66 &    8  &   19 &  08 10 13.53 & -49 20 44.05 & 1114.921 &  585.299 &    10.970   &   0.022  & 00 &   0.478  &    0.027 &  00  &  0.606  &    0.030 &  00 \\
  67 &   19  &   14 &  08 11 11.36 & -49 04 44.47 & 1276.119 & 1332.991 &    11.051   &   0.022  & 00 &   0.327  &    0.027 &  00  &  0.447  &    0.030 &  00 \\
\end{tabular}
\end{table*}
\end{landscape}
\newpage
\begin{landscape}
\textwidth=240mm
\begin{table*}
  \begin{tabular}{@{}ccccc..cccccccccc@{}}
            &       &      & 
\multicolumn{2}{c}{J2000 Position} & 
\multicolumn{2}{c}{Pixel Position} & 
\multicolumn{3}{c}{V} &
\multicolumn{3}{c}{B-V} &
\multicolumn{3}{c}{V-I} \\
Clari\'a ID & Field & Star &                  
R.A. & Dec. &                
X & Y &
Mag. & Error & Flag  & 
Mag. & Error & Flag  & 
Mag. & Error & Flag  \\ 
  68 &   18  &   23 &  08 10  0.89 & -49 08 32.58 & 1426.108 & 1902.766 &    10.983   &   0.022  & 00 &   0.452  &    0.027 &  00  &  0.580  &    0.030 &  00 \\
  69 &   13  &   38 &  08 10 27.14 & -49 10 29.45 &  776.268 &  628.323 &    11.450   &   0.022  & 04 &   0.878  &    0.024 &  04  &  1.361  &    0.026 &  40 \\
  70 &   12  &   16 &  08 11 22.34 & -49 09 14.84 &  996.397 &  440.142 &    11.137   &   0.022  & 00 &   1.355  &    0.027 &  00  &  1.424  &    0.030 &  00 \\
  71 &   14  &   22 &  08 08 50.38 & -49 16 16.43 & 1562.658 & 1492.286 &    11.382   &   0.022  & 00 &   0.422  &    0.027 &  00  &  0.557  &    0.030 &  00 \\
  72 &   17  &   17 &  08 09 14.03 & -49 04  3.30 & 1000.210 & 1229.880 &    11.412   &   0.022  & 00 &   0.366  &    0.027 &  00  &  0.420  &    0.030 &  00 \\
  73 &    8  &   21 &  08 09 47.31 & -49 19 42.72 & 1754.214 &  434.597 &    11.435   &   0.022  & 00 &   1.322  &    0.024 &  00  &  1.374  &    0.030 &  00 \\
  74 &   14  &   21 &  08 09  2.85 & -49 16  9.65 & 1258.535 & 1474.364 &    11.460   &   0.022  & 00 &   0.211  &    0.027 &  00  &  0.219  &    0.031 &  00 \\
  75 &    9  &   15 &  08 10 40.05 & -49 17 47.30 & 2035.372 &  148.636 &    11.565   &   0.016  & 00 &   1.187  &    0.018 &  00  &  1.202  &    0.021 &  00 \\
  76 &    8  &   22 &  08 10 11.66 & -49 22 27.71 & 1159.697 &  843.840 &    11.601   &   0.022  & 00 &   0.556  &    0.027 &  00  &  0.635  &    0.030 &  00 \\
  77 &    8  &   24 &  08 10 16.64 & -49 21 32.72 & 1038.872 &  706.433 &    11.625   &   0.022  & 00 &   1.772  &    0.024 &  00  &  1.929  &    0.030 &  04 \\
  78 &    3  &   25 &  08 11  5.62 & -49 03 19.43 &  405.379 &  381.260 &    11.705   &   0.005  & 00 &   0.977  &    0.006 &  00  &  1.087  &    0.030 &  00 \\
  79 &    3  &   24 &  08 10 39.83 & -49 04 38.04 & 1036.966 &  578.271 &    11.604   &   0.022  & 00 &   0.465  &    0.027 &  00  &  0.614  &    0.008 &  00 \\
  80 &   18  &   29 &  08 10 18.37 & -49 06 46.32 &  999.032 & 1636.669 &    11.870   &   0.005  & 00 &   0.536  &    0.006 &  00  &  0.664  &    0.006 &  00 \\
  81 &    3  &   26 &  08 10 35.30 & -49 06 28.31 & 1147.143 &  853.403 &    11.808   &   0.022  & 00 &   0.242  &    0.027 &  00  &  0.292  &    0.006 &  00 \\
  83 &    3  &   27 &  08 10 49.85 & -49 11 25.84 &  789.433 & 1594.123 &    11.799   &   0.005  & 00 &   0.466  &    0.006 &  00  &  0.593  &    0.006 &  00 \\
  84 &   18  &   27 &  08 10  1.44 & -48 58 25.49 & 1417.666 &  389.411 &    11.881   &   0.022  & 00 &   0.212  &    0.027 &  00  &  0.272  &    0.031 &  00 \\
  85 &    8  &   28 &  08 10 25.25 & -49 18 58.77 &  829.951 &  322.177 &    12.031   &   0.022  & 00 &   0.563  &    0.028 &  00  &  0.709  &    0.031 &  00 \\
  86 &    3  &   29 &  08 11 17.43 & -49 06 15.21 &  115.758 &  819.153 &    12.021   &   0.003  & 00 &   0.976  &    0.004 &  00  &  1.098  &    0.005 &  00 \\
  87 &    1  &   32 &  08 11 20.74 & -49 13 10.18 &   33.468 &  281.191 &    12.016   &   0.002  & 00 &   0.536  &    0.003 &  00  &  0.650  &    0.003 &  00 \\
  88 &   12  &   24 &  08 11 28.86 & -49 10 52.64 &  836.330 &  683.557 &    12.103   &   0.022  & 00 &   0.445  &    0.028 &  00  &  0.515  &    0.026 &  00 \\
  89 &   18  &   30 &  08 10 34.32 & -49 00 49.89 &  610.018 &  747.421 &    11.985   &   0.022  & 00 &   0.444  &    0.027 &  00  &  0.567  &    0.031 &  00 \\
  90 &    3  &   35 &  08 10 29.53 & -49 09  0.68 & 1287.120 & 1233.552 &    12.173   &   0.005  & 00 &   1.400  &    0.006 &  00  &  1.434  &    0.008 &  04 \\
  91 &    3  &   36 &  08 10 20.14 & -49 09 38.66 & 1516.448 & 1328.978 &    12.157   &   0.004  & 00 &   1.342  &    0.005 &  00  &  1.312  &    0.007 &  00 \\
  92 &   18  &   32 &  08 10 23.04 & -49 00 30.32 &  886.729 &  699.135 &    12.098   &   0.022  & 00 &   0.203  &    0.027 &  00  &  0.318  &    0.026 &  00 \\
  93 &    3  &   37 &  08 11 11.59 & -49 05 21.17 &  258.985 &  684.537 &    12.190   &   0.004  & 00 &   1.747  &    0.005 &  00  &  1.937  &    0.030 &  00 \\
  94 &   18  &   35 &  08 10 45.45 & -49 01  7.05 &  336.790 &  789.856 &    12.108   &   0.005  & 00 &   0.497  &    0.006 &  00  &  0.662  &    0.006 &  00 \\
  95 &    3  &   38 &  08 10 27.13 & -49 11 50.45 & 1344.498 & 1656.863 &    12.267   &   0.003  & 00 &   1.078  &    0.004 &  00  &  1.192  &    0.005 &  00 \\
  96 &    1  &   36 &  08 10 57.89 & -49 18 15.75 &  590.523 & 1043.333 &    12.257   &   0.004  & 00 &   1.583  &    0.005 &  00  &  1.706  &    0.018 &  00 \\
  97 &   14  &   32 &  08 09 47.32 & -49 13  5.35 &  173.533 & 1012.938 &    12.301   &   0.016  & 00 &   0.536  &    0.017 &  00  &  0.698  &    0.019 &  00 \\
  98 &    1  &   38 &  08 11  0.67 & -49 15 15.04 &  523.348 &  592.822 &    12.338   &   0.005  & 00 &   1.507  &    0.006 &  00  &  1.536  &    0.026 &  00 \\
  99 &    3  &   39 &  08 10 34.72 & -49 08 40.21 & 1160.424 & 1182.160 &    12.385   &   0.003  & 00 &   0.561  &    0.004 &  00  &  0.686  &    0.005 &  00 \\
 100 &    8  &   35 &  08 10  9.29 & -49 20 54.14 & 1218.054 &  610.737 &    12.449   &   0.015  & 00 &   0.586  &    0.019 &  00  &  0.771  &    0.022 &  00 \\
 101 &    3  &   48 &  08 10 56.58 & -49 09  0.59 &  625.518 & 1231.824 &    12.535   &   0.003  & 00 &   0.386  &    0.004 &  00  &  0.525  &    0.005 &  00 \\
 102 &    2  &   40 &  08 10 12.23 & -49 13  5.83 &  147.023 &  263.520 &    12.500   &   0.002  & 00 &   0.565  &    0.002 &  00  &  0.690  &    0.002 &  00 \\
 103 &    3  &   53 &  08 10 58.55 & -49 08 26.04 &  577.501 & 1145.637 &    12.764   &   0.003  & 00 &   1.027  &    0.004 &  00  &  1.147  &    0.005 &  00 \\
 104 &   18  &   46 &  08 10 24.87 & -49 01 49.74 &  841.404 &  897.015 &    12.664   &   0.003  & 00 &   0.605  &    0.004 &  00  &  0.730  &    0.005 &  00 \\
 105 &    2  &   33 &  08 10  1.67 & -49 22  6.38 &  403.478 & 1611.086 &    13.086   &   0.003  & 00 &   1.342  &    0.004 &  00  &  1.361  &    0.005 &  00 \\
 106 &    1  &   52 &  08 10 33.81 & -49 11 24.90 & 1180.583 &   20.619 &    12.815   &   0.002  & 00 &   1.116  &    0.003 &  00  &  1.208  &    0.004 &  00 \\
 107 &    1  &   53 &  08 11  5.23 & -49 13 40.42 &  412.225 &  356.847 &    12.938   &   0.002  & 00 &   0.424  &    0.003 &  00  &  0.559  &    0.003 &  00 \\
 108 &    3  &   71 &  08 10 24.55 & -49 08 48.84 & 1409.036 & 1204.441 &    13.190   &   0.003  & 00 &   0.544  &    0.004 &  00  &  0.686  &    0.005 &  00 \\
 109 &    1  &   61 &  08 10 48.17 & -49 23 38.66 &  825.958 & 1848.613 &    13.173   &   0.003  & 00 &   0.690  &    0.004 &  00  &  0.782  &    0.005 &  00 \\
 110 &   13  &   93 &  08 10 15.42 & -49 11  9.77 & 1062.406 &  729.484 &    13.209   &   0.002  & 00 &   0.756  &    0.003 &  00  &  0.892  &    0.004 &  00 \\
 111 &    3  &   93 &  08 11  0.11 & -49 06 44.44 &  539.878 &  892.355 &    13.289   &   0.003  & 00 &   0.708  &    0.004 &  00  &  0.810  &    0.005 &  00 \\
 112 &    3  &   73 &  08 10 29.37 & -49 08  2.56 & 1291.456 & 1088.710 &    13.310   &   0.003  & 00 &   0.380  &    0.004 &  00  &  0.481  &    0.005 &  00 \\
 113 &   13  &  148 &  08 10 27.05 & -49 08 49.05 &  778.787 &  378.041 &    13.239   &   0.003  & 00 &   0.697  &    0.004 &  00  &  0.830  &    0.005 &  00 \\
 114 &   13  &  300 &  08 10 25.77 & -49 08 12.73 &  810.332 &  287.572 &    13.536   &   0.003  & 00 &   0.964  &    0.004 &  00  &  1.167  &    0.005 &  00 \\
 115 &    3  &   94 &  08 10 33.24 & -49 07 38.72 & 1196.981 & 1029.006 &    13.420   &   0.003  & 00 &   1.382  &    0.004 &  00  &  1.444  &    0.005 &  00 \\
 116 &    3  &   88 &  08 10 34.84 & -49 08 17.60 & 1157.514 & 1125.817 &    13.474   &   0.003  & 00 &   0.481  &    0.004 &  00  &  0.640  &    0.005 &  00 \\
 117 &    1  &   83 &  08 11  3.48 & -49 13 26.53 &  455.192 &  322.276 &    13.537   &   0.002  & 00 &   0.567  &    0.003 &  00  &  0.715  &    0.003 &  00 \\
 118 &    1  &  166 &  08 10 40.37 & -49 21 50.95 & 1016.383 & 1580.579 &    13.948   &   0.003  & 00 &   0.627  &    0.004 &  00  &  0.764  &    0.005 &  00 \\

\end{tabular}
\vskip 5mm
See Sections \ref{optphot} and \ref{comb_meas} for 
1=non-stellar; 2=too close to detector edge; 3=fit to sky histogram failed; 
4=saturated; 8
=bad pixel; 9=negative counts.
\end{table*}
\end{landscape}

\bsp

\label{lastpage}

\end{document}